\documentclass[aps, pre, twocolumn, groupedaddress, superscriptaddress, longbibliography]{revtex4-1}

\usepackage{graphicx}
\usepackage{color}
\usepackage{amsmath}
\usepackage{amsfonts}
\usepackage{amssymb}
\usepackage{dcolumn}
\usepackage{hyperref}
\usepackage[justification=centerlast]{caption}
\hypersetup{colorlinks=true, urlcolor=blue, linkcolor=blue, citecolor=blue}
\usepackage[draft]{todonotes}   

\newcommand*\dif{\mathop{}\!\mathrm{d}}

\begin{document}
\title{Turning intractable counting into sampling: computing the
  configurational entropy of three-dimensional jammed packings}
\author{Stefano Martiniani}
\email{sm958@cam.ac.uk}
\affiliation{Department of Chemistry, University of Cambridge, Lensfield Road, Cambridge, CB2 1EW, UK}
\author{K. Julian Schrenk}
\email{kjs73@cam.ac.uk}
\affiliation{Department of Chemistry, University of Cambridge, Lensfield Road, Cambridge, CB2 1EW, UK}
\author{Jacob D. Stevenson}
\affiliation{Microsoft Research Ltd, 21 Station Road, Cambridge, CB1 2FB, UK}
\affiliation{Department of Chemistry, University of Cambridge, Lensfield Road, Cambridge, CB2 1EW, UK}
\author{David J. Wales}
\affiliation{Department of Chemistry, University of Cambridge, Lensfield Road, Cambridge, CB2 1EW, UK}
\author{Daan Frenkel}
\affiliation{Department of Chemistry, University of Cambridge, Lensfield Road, Cambridge, CB2 1EW, UK}

\begin{abstract}
We report a numerical calculation of the total number of disordered
jammed configurations $\Omega$ of $N$ repulsive, three-dimensional
spheres in a fixed volume $V$.  To make these calculations tractable,
we increase the computational efficiency of the approach of Xu
\emph{et al.} (Phys. Rev. Lett. {\bf 106}, 245502 (2011)) and Asenjo
\emph{et al.} (Phys. Rev. Lett. {\bf 112}, 098002 (2014)) and we
extend the method to allow computation of the configurational entropy
as a function of pressure. The approach that we use computes the
configurational entropy by sampling the absolute volume of basins of
attraction of the stable packings in the potential energy landscape.
We find a surprisingly strong correlation between the pressure of a
configuration and the volume of its basin of attraction in the
potential energy landscape. This relation is well described by a power
law.  Our methodology to compute the number of minima in the potential
energy landscape should be applicable to a wide range of other
enumeration problems in statistical physics, string theory, cosmology
and machine learning, that aim to find the distribution of the extrema
of a scalar cost function that depends on many degrees of freedom.
\end{abstract}
\maketitle
\makeatletter{}\section{Introduction}
Many questions in physics are easy to pose but difficult to answer.
One such question is: how many microscopic states of a given system
are compatible with its macroscopic properties?  In statistical
mechanics, knowledge of this number allows us to compute the entropy,
and thereby predict the macroscopic properties of a system from
knowledge of the interaction between atoms or molecules.

In granular matter we can similarly ask how many microstates are
compatible with a given set of macroscopic properties.  However, the
computation of the corresponding absolute entropy has thus far proven
to be extremely challenging.  Without such knowledge, it is not
possible to explore the analogies and differences between granular and
Boltzmann entropy. Being able to compute the configurational entropy
is therefore clearly important. The more so as granular materials are
ubiquitous in everyday life (sand, soil, powders).  Many industrial
processes involve granular materials. In the natural world, the
Earth's surface contains vast granular assemblies such as dunes, which
interact with wind, water, and vegetation \cite{Kok12}. Packings of
particles that are soft or biological in nature, such as cells,
hydrogels and foams are also known to undergo jamming
\cite{van2010jamming} and their behaviour to be ``granular''
\emph{viz} not subject to thermal motion. Moreover, as glasses and
granular materials share many properties it has been proposed that
their physics may be controlled by the same underlying principles
\cite{Liu98}.

The study of granular materials is complicated by the fact that these
materials are intrinsically out-of-equilibrium. In fact, thermal
motion plays no role in granular matter. It maintains its
configuration unless driven by external forces. As a consequence, the
properties of granular materials depend upon their preparation
protocol.

Granular materials are athermal and cannot therefore be described by
statistical mechanics.  However, these materials can exists in a very
large number of distinct states and this fact inspired Edwards and
Oakeshott \cite{Edwards89} well over two decades ago to propose a
statistical-mechanics-like formalism to describe the properties of
granular matter.  In its original version, the Edwards theory assumed
that all mechanically stable configurations (`jammed' states) are
equally probable and that the logarithm of the number of these states
plays a role similar to that of entropy.  In this theoretical
framework, the volume of the system and its \emph{compactivity}
(i.e. the derivative of volume with respect to the configurational
entropy) are the analogues of the energy and temperature in thermal
systems.

In the absence of explicit calculations (or measurements) of the
absolute configurational entropy, a direct test of the Edwards
hypothesis has proven difficult, and different authors have arrived at
different conclusions based on indirect tests in either
simulations~\cite{Makse02, Song05, Barrat01, Gao06} or
experiments~\cite{Lechenault06, Daniels08}. In addition, alternative
definitions of entropy have been proposed to characterise the
complexity of granular systems while circumventing explicit
enumeration of states \cite{Brujic07, Baranau13}.

Numerous tests of the Edwards volume ensemble have focused on the
determination of the compactivity~\cite{Nowak98, Schroter05,
  Mcnamara09, Dean03, Zhao12, Aste08, Blumenfeld03, Blumenfeld12,
  Zhao14}. However, the role of compactivity as a temperature-like
quantity is problematic as Puckett and Daniels \cite{Puckett13,Bi15}
have shown that it does not satisfy the equivalent of the zero-th law
of Thermodynamics - the law that is the basis of all thermometry.

Edwards' theory has been generalised to include the distribution of
stresses within the system through the force-moment tensor
\cite{Briscoe08, Blumenfeld09, Henkes07, Henkes09} and another
analogue of temperature emerged, known as \emph{angoricity}, which is
a measure of the change in entropy with stress.  The experiments by
Puckett and Daniels \cite{Puckett13} showed that angoricity, unlike
compactivity, is a temperature-like quantity as it satisfies the
zero-th law.

To date only a few examples of numerical tests of the generalised
Edwards ensemble are available \cite{Henkes07, Wang10, Wang12,
  Puckett13}. Numerical tests of the stress ensemble focus on systems
of soft spheres near jamming where the compactivity $X \to \infty$ and
fluctuations in volume are negligible compared to stress fluctuations
\cite{Henkes07, Henkes09}. Wang \emph{et al.} \cite{Wang10, Wang12}
proposed a unified test that compared ensemble averaged results over
volume and stress with predictions for the jamming transition, finding
agreement; we note, however, that in the latter approach the results
rely significantly on the equiprobability assumption.

When the system is composed of very stiff grains, or is close to
jamming, any small deformation will lead to a large change in the
contact forces. In these limits the geometric and the force degrees of
freedom can be decoupled, giving rise to the force network ensemble
\cite{Snoeijer04} (FNE). In this framework, force networks are
constructed on a given geometry and each force state is assumed to be
equiprobable. The FNE has been utilised as a testing ground for
statistical frameworks \cite{Van09, Tighe10, Tighe11}.

More than two decades after its introduction many fundamental
questions concerning the Edwards hypothesis remain unanswered.  This
unsatisfactory state of affairs is at least partly due to the fact
that no efficient methods existed to measure or compute the absolute
configurational entropy directly. Until recently, the only way to
determine the configurational entropy was by direct enumeration of the
distinct jammed states of a system. This method is inefficient and
cannot be used for systems that contain more than 10-20
particles. Over the past few years, the situation on the numerical
front has changed: recent numerical work by Asenjo \emph{et al.}
\cite{Asenjo14,Paillusson15}, based on an approach introduced by Xu
\emph{et al.}  \cite{Xu11}, has demonstrated that it is possible to
compute the number of distinct jammed states of a system, even when
this number is far too large (e.g. 10$^{250}$) to allow direct
enumeration. The approach of Refs.~\cite{Xu11, Asenjo14, Paillusson15}
replaces an intractable enumeration problem by a tractable scheme to
sample the (absolute) volume of the basins of attraction of stable
states in the potential energy landscape.

The approach described herein is completely general and it extends to
any energy landscape problem that aims to find the extrema of a scalar
cost function that depends on many degrees of freedom. Enumerating the
number of solutions or stationary points, and their distribution, for
certain classes of random functions is a classical problem in
mathematics and statistics \cite{kac1948average, farahmand1986average,
  bogomolny1992distribution, edelman1995many, rojas1996average,
  kostlan2000expected, malajovich2004high, azais2005roots,
  armentano2009random, fyodorov2012freezing, nicolaescu2012complexity,
  fyodorov2013high, cheng2015explicit, fyodorov2015number}. In
statistical physics, \emph{ad hoc} numerical and theoretical methods
have been developed in the realms of random Gaussian and polynomial
fields \cite{fyodorov2004complexity, bray2007statistics,
  fyodorov2012critical, auffinger2013random, fyodorov2014topology,
  hughes2014inversion, mehta2014communication, cleveland2014certified,
  mehta2015statistics}. In this sense, particular attention has been
devoted to the mean-field p-spin spherical model
of a spin glass with quenched disorder
\cite{cugliandolo1993analytical, kurchan1993barriers, mehta2013energy,
  auffinger2013complexity, subag2015complexity}. A related area is
the computation of the configurational contribution to the entropy of
structural glasses \cite{sastry2001relationship,
  banerjee2014role}. The physical significance of this method goes
even further to encompass string theory \cite{distler2005random,
  douglas2007flux, greene2013tumbling}, cosmology
\cite{aazami2006cosmology, tye2009multi, frazer2011exploring,
  frazer2012multi, battefeld2012unlikeliness} and machine learning
\cite{wainrib2013topological, choromanska2014loss,
  sagun2014explorations, chaudhari2015trivializing}.

We note that the geometrical structure of the basins of attraction of
jammed states had been studied by O'Hern and co-workers \cite{OHern03,
  Gao06, Ashwin12}.  O'Hern also reported direct enumeration estimates
of the number of jammed states of small systems. A rather different
technique (`basin sampling') to count the number of energy minima in
the potential energy landscape of small clusters had been reported by
Wales and co-workers ~\cite{Wales03, Ballard15}.

We note that, for the system (and protocol) considered by Asenjo
\emph{et al.}, not all packings are equally probable.  However, as
shown in Ref.~\cite{Asenjo14}, the equal-probability hypothesis is not
needed to arrive at a meaningful definition of an extensive granular
entropy.  When, in the remainder of the present paper, we mention the
configurational (granular) entropy, we refer to the definition of
Ref.~\cite{Asenjo14}.  

We stress that, even though the approach of Refs.~\cite{Xu11,
  Asenjo14, Paillusson15} allows to solve enumeration problems which
were far from possible using direct enumeration, it is still
computationally expensive.  Thus far, it had only been applied to
two-dimensional packings.  Substantial `technical' improvements were
needed to make the method fast enough to deal with three-dimensional
systems.  

\begin{figure}
\includegraphics[width=\columnwidth]{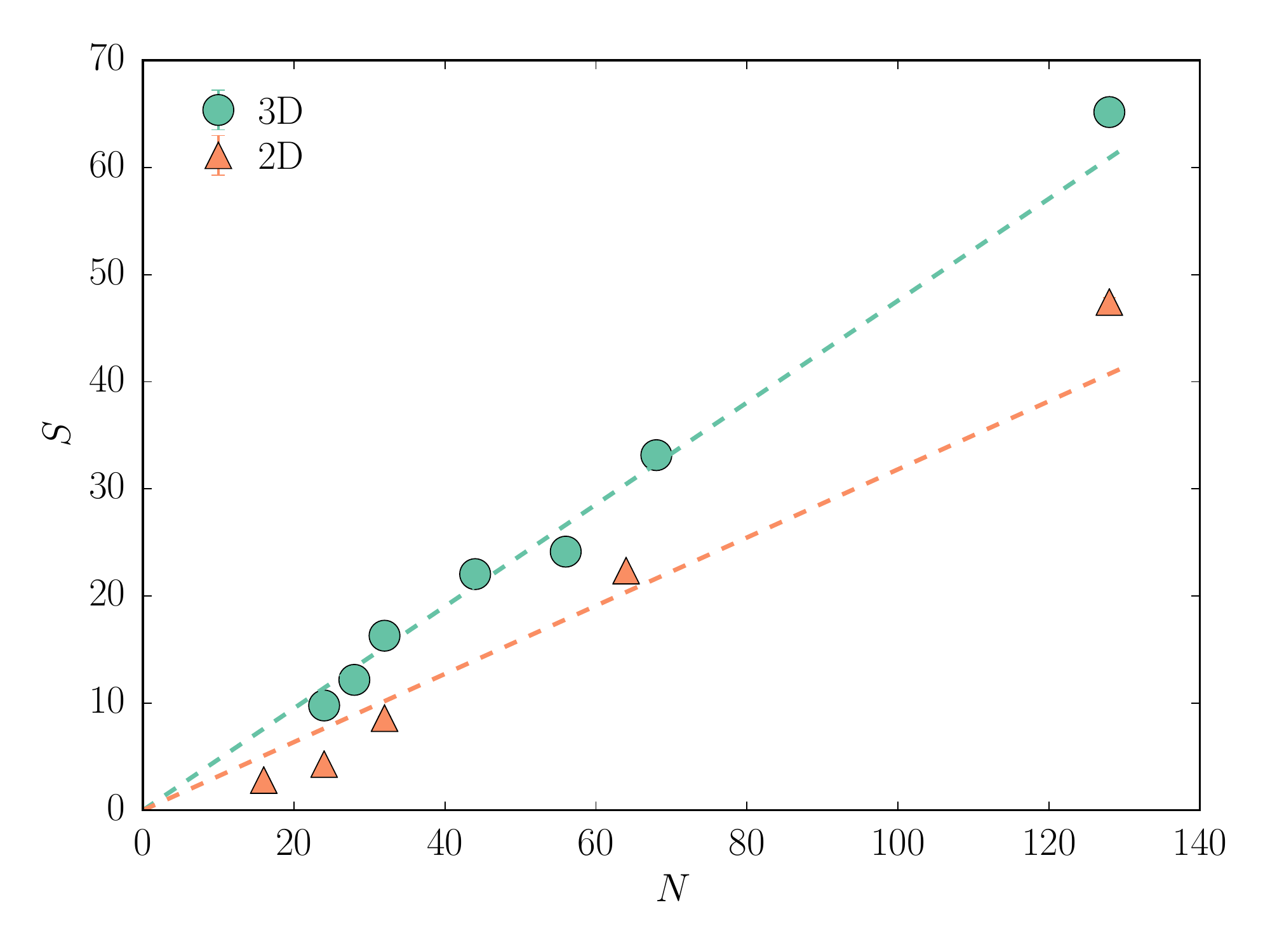}
  \caption{\label{fig::compare_entropy} Entropy as a function of
    system size $N$ for two (Ref.~\cite{Asenjo14}) and
    three-dimensional (this work) jammed sphere packings. Dashed
    curves are lines of best fit of the form $S=aN$.}
\end{figure}

In the present paper, we present the first enumeration of the number
of jammed packings for three-dimensional systems consisting of up to
$128$ soft spheres. A direct comparison of the entropy measured as a
function of system size for two and three-dimensional jammed sphere
packings is shown in Fig.~\ref{fig::compare_entropy}. The potential of
the method presented herein can be verified unequivocally from
Fig.~\ref{fig::compare_entropy}: we are able of tackling problems at
least $500$ million times more complex, and of greater computational
cost, than the already spectacularly difficult questions confronted by
Asenjo et al. \cite{Asenjo14}. Furthermore we show how our improved
procedure allows first-principles computation of configurational
entropy as a function of system size and pressure. The method and the
technical improvements needed to overcome this numerical challenge are
presented alongside the main results.

The remainder of this work is organised as follows.
Section~\ref{sec::basic_principle} describes the basic principle of
the mean basin volume method for counting, and explains how that
strategy can be applied to enumerate granular packings. The
enumeration and entropy results for three-dimensional jammed sphere
packings as a function of system size and pressure are reported in
Sec.~\ref{sec::r3d}. The rest of the manuscript is dedicated to the
description of the improved numerical
method. Section~\ref{sec::packing_sampling_protocol} outlines our
protocol for sampling different granular packings, and it describes
the corresponding potential energy landscape and minimisation
techniques.  Application of thermodynamic integration to compute the
volume of a basin of attraction in such a landscape is described in
Sec.~\ref{sec::basin_volume_thermodynamic_integration}.  Aspects of
the data analysis tools used on the histograms of sampled basin
volumes, and related configurational entropy definitions, are described in
Sec.~\ref{sec::volume_distributions_data_analysis}.  Conclusions are
drawn in Sec.~\ref{sec::conclusion}.  Further technical background is
given in the appendices.
 
\makeatletter{}\begin{figure*}
\centering
\begin{minipage}[b]{.45\textwidth}
\includegraphics[width=\columnwidth]{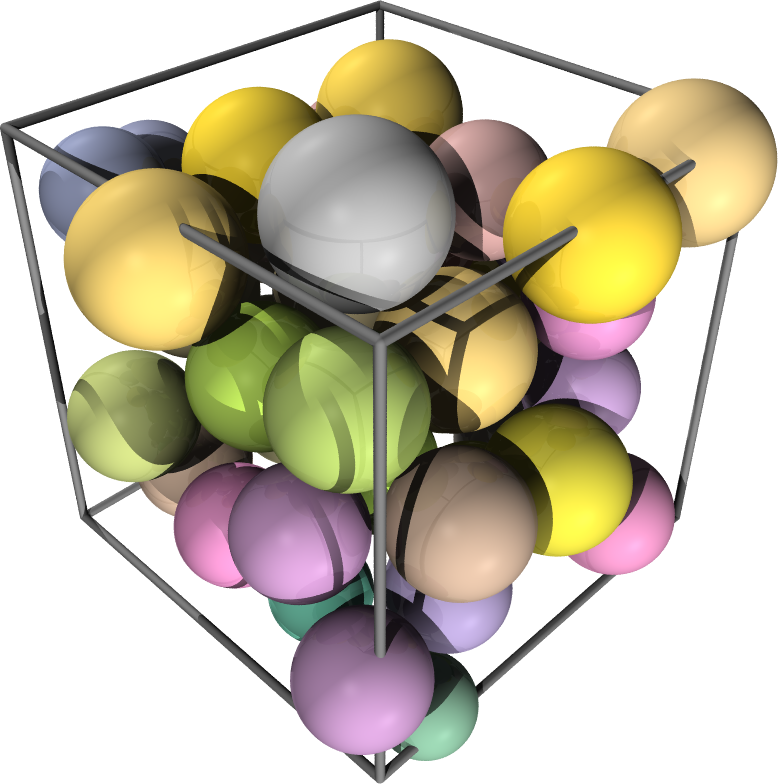}
\end{minipage}\qquad
\begin{minipage}[b]{.45\textwidth}
\includegraphics[width=\columnwidth]{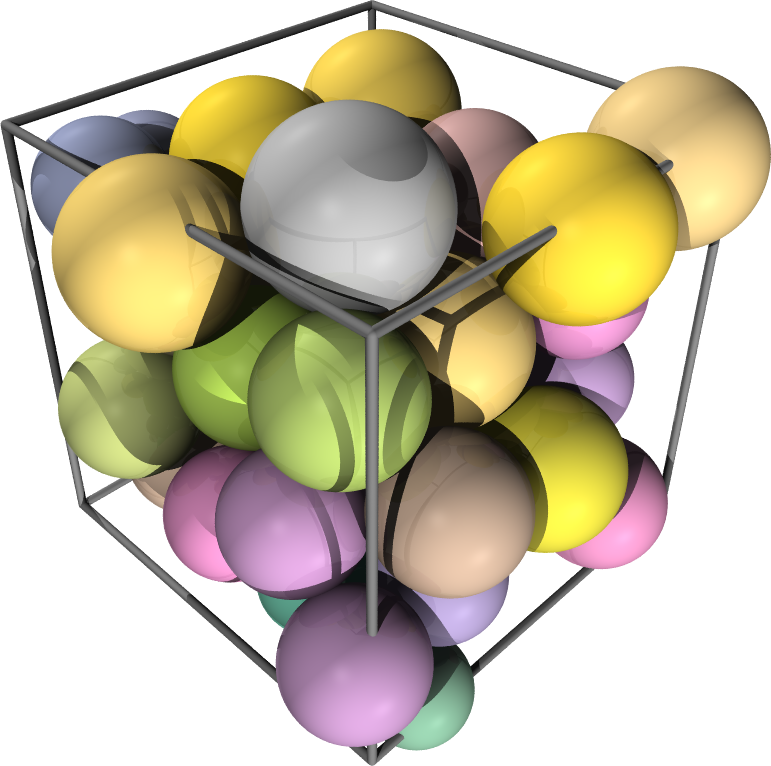}
\end{minipage}
\caption{\label{fig:3d_packings} Hard sphere fluid at $\phi_\text{HS}
  = 0.5$, left, and HS-WCA jammed packing at $\phi_\text{SS} = 0.7$,
  right, for a system of $44$ polydisperse hard spheres with mean
  radius $\langle r_h \rangle=1$ and standard deviation
  $\sigma_\text{HS}=0.05$.  We prepare the polydisperse HS fluid
  configurations at fixed packing fraction $\phi_{\mathrm{HS}}=0.5$ by
  a Monte Carlo simulation. Particles are then inflated by the same
  factor, proportional to their radius (spheres are coloured according
  to their radius), to obtain an over-compressed soft spheres jammed
  packing at $\phi_{\mathrm{SS}}=0.7$ by an infinitely fast quench
  (energy minimisation).}
\end{figure*}

\section{\label{sec::basic_principle}Basic principle: counting by
  sampling}
In this section, we briefly review the numerical approach that we use
to compute the number of distinct jammed states.  We stress that the
approach that we use has much wider applicability than the counting of
granular packings~\cite{fyodorov2004complexity, bray2007statistics,
  fyodorov2012critical, auffinger2013random, fyodorov2014topology,
  hughes2014inversion, mehta2014communication, cleveland2014certified,
  mehta2015statistics, cugliandolo1993analytical, kurchan1993barriers,
  mehta2013energy, auffinger2013complexity, subag2015complexity,
  sastry2001relationship, banerjee2014role, distler2005random,
  douglas2007flux, aazami2006cosmology, tye2009multi,
  frazer2011exploring, frazer2012multi, battefeld2012unlikeliness,
  wainrib2013topological, choromanska2014loss, sagun2014explorations,
  chaudhari2015trivializing, Frenkel13}.  In the context of granular
packings, our aim is to compute the number of ways $\Omega$ in which
$N$ spheres can be arranged in a given volume $V_\text{box}$ of
Euclidean dimension $d$.  Knowledge of $\Omega$ allows us to compute
configurational entropies and related quantities from first principles
\cite{Edwards89, Asenjo14}.  Our approach is based on a rigorous
mapping of the enumeration problem onto counting the number of minima
of a potential energy landscape \cite{Xu11}. The approach makes no use
of a harmonic~\cite{Speedy99} or quasi-harmonic \cite{Wales13}
approximation.  For a system of hard particles the potential energy
function is discontinuous, that is, the energy of the system is either
zero, if no two particles overlap, or infinity otherwise.  Then, at
jamming, in the absence of rattlers, basins of attraction are single
points in configuration space and they have no associated volume.
This does not mean that we cannot sample the energy minima of a system
of hard particles. The reason is that all jammed structures of hard
particles correspond to the zero potential energy minima of a system
with a continuous repulsive potential with the same range as the
hard-core diameter of the hard particles. In what follows, we focus on
this class of systems, but we generalise the problem by also
considering minima with a non-zero potential energy. In particular, we
consider spherical particles with a hard core and a short-ranged
continuous repulsive interaction. Under conditions where this system
is jammed, a system with {\em only} the hard-core interactions would
still be fluid and would sample the accessible configuration space
uniformly. This remaining accessible volume is partitioned in basins
of attraction defined by the soft shells. The HS-WCA potential used to
simulate hard-core plus soft-shell interactions and the packing
preparation protocol are described in Sec~\ref{sec:wca_pot}. For an
illustration of the packing preparation protocol refer to
Fig.~\ref{fig:3d_packings}. As we argue below, using an HS-WCA model
greatly improves the efficiency of determination of basin volumes.

Let us denote the total available volume in $dN$-dimensional space as
$\mathcal{V}$.  Note that $\mathcal{V}$ is not the total volume of
configuration space ($V^N$), but just that part of the volume that is
free of hard-core overlaps. It is the configurational part of the
partition function of the hard-core system at the number density under
consideration. Since the accessible configuration space is tiled by
the basins of attraction of the distinct energy minima
\cite{Stillinger84, Stillinger85, Stillinger95, Speedy99} we can
write:
\begin{equation}
    \mathcal{V} = \sum_{i=1}^{\Omega} v_i,
\end{equation}
where $v_i$ is the volume of the $i$-th basin of attraction and
$\Omega$ is the total number of distinct minima.  We thus make the
simple observation:
\begin{equation}
    \sum_{i=1}^{\Omega} v_i = \frac{\Omega}{\Omega}\sum_{i=1}^{\Omega}
    v_i = \Omega \langle v \rangle,
\end{equation}
where $\langle v \rangle$ is the \emph{mean basin volume}, from which
it follows immediately that
\begin{equation}
    \label{eq:bv_omega}
    \Omega = \frac{\mathcal{V}}{\langle v \rangle}.
\end{equation}
We note that, for sphere packings, $\mathcal{V}$ is known from the
equation of state of the underlying hard sphere fluid (see
Appendix~\ref{sec::poly_fluid}) and we can measure $\langle v \rangle$
by thermodynamic integration, as discussed in detail in the
Sec~\ref{sec::packing_sampling_protocol}.  The approach of~\cite{Xu11,
  Asenjo14} has thus turned the intractable enumeration problem of
finding $\Omega$ into a sampling one, namely measuring $\langle v
\rangle$.
 
\makeatletter{}\section{\label{sec::r3d}Results: counting disordered 3D sphere packings}

The mean basin volume method for enumerating the number of
mechanically stable packings was introduced by Xu \emph{et al.}
\cite{Xu11}, and tested on a small system of soft disks.  Asenjo
\emph{et al.}  \cite{Asenjo14} then made a number of modifications to
the algorithm that allowed them to apply it to larger systems of up to
128 disks. As was the case with Ref.~\cite{Xu11}, the calculations of
Ref.~\cite{Asenjo14} focused on two-dimensional systems because of the
high computational costs involved in studying 3D systems.  Here we
present results for systems of three-dimensional soft spheres. We are
thus in a position to compute the configurational entropy of a real
(but idealised) three-dimensional system.

We first describe an analysis similar to the one reported by Asenjo
\emph{et al.} \cite{Asenjo14} to verify the extensivity of the entropy
$S(V)$ at constant packing fraction. Next, we extend our approach to
the generalised Edwards ensemble, i.e.~one based on a description
of the system in terms of its volume and
pressure, to compute the generalised entropy $S(V, \mathcal{P})$.

We investigate three-dimensional packings with system sizes ranging
from $24$ to $128$ HS-WCA particles, see Eq.~(\ref{eq:hswca}), at
$\phi_\text{HS}=0.5$ hard-sphere fluid packing fraction and
$\phi_\text{SS}=0.7$ soft sphere packing fraction, corresponding to a
ratio of the soft and hard-sphere radii ratio $r_\text{SS}/r_\text{HS}
= 1.12$, prepared following the protocol outlined in
Sec.~\ref{sec::packing_sampling_protocol}. For each system size we
compute the volume of the basin of attraction of approximately $1000$
packings.  Each PT run (see Sec.~\ref{sec::basin_volume_thermodynamic_integration}) was performed on $15$ parallel threads of a
single eight-core dual Xeon E$5-2670$ ($2.6$GHz, Westmere) node. Our
choice of convergence criterion was such that when the uncorrelated
statistical error for each of the replicas' mean square displacement
fell below $5\%$ the calculations were terminated. This set-up
translated in run times ranging from $10$ to $300$ hours per basin
depending on system size, which amounts to $\mathcal{O}(10^6)$ hours
of total run time and $\mathcal{O}(10^7)$ total cpu hours.  We then
analyse the corresponding distribution of dimensionless free energies
following the protocol described in
Sec.~\ref{sec::basin_volume_thermodynamic_integration} and
\ref{sec::volume_distributions_data_analysis} and summarised in
Appendix~\ref{sec::method_summary}.

\subsection{Extensivity of the entropy}
\begin{figure}
\includegraphics[width=\columnwidth]{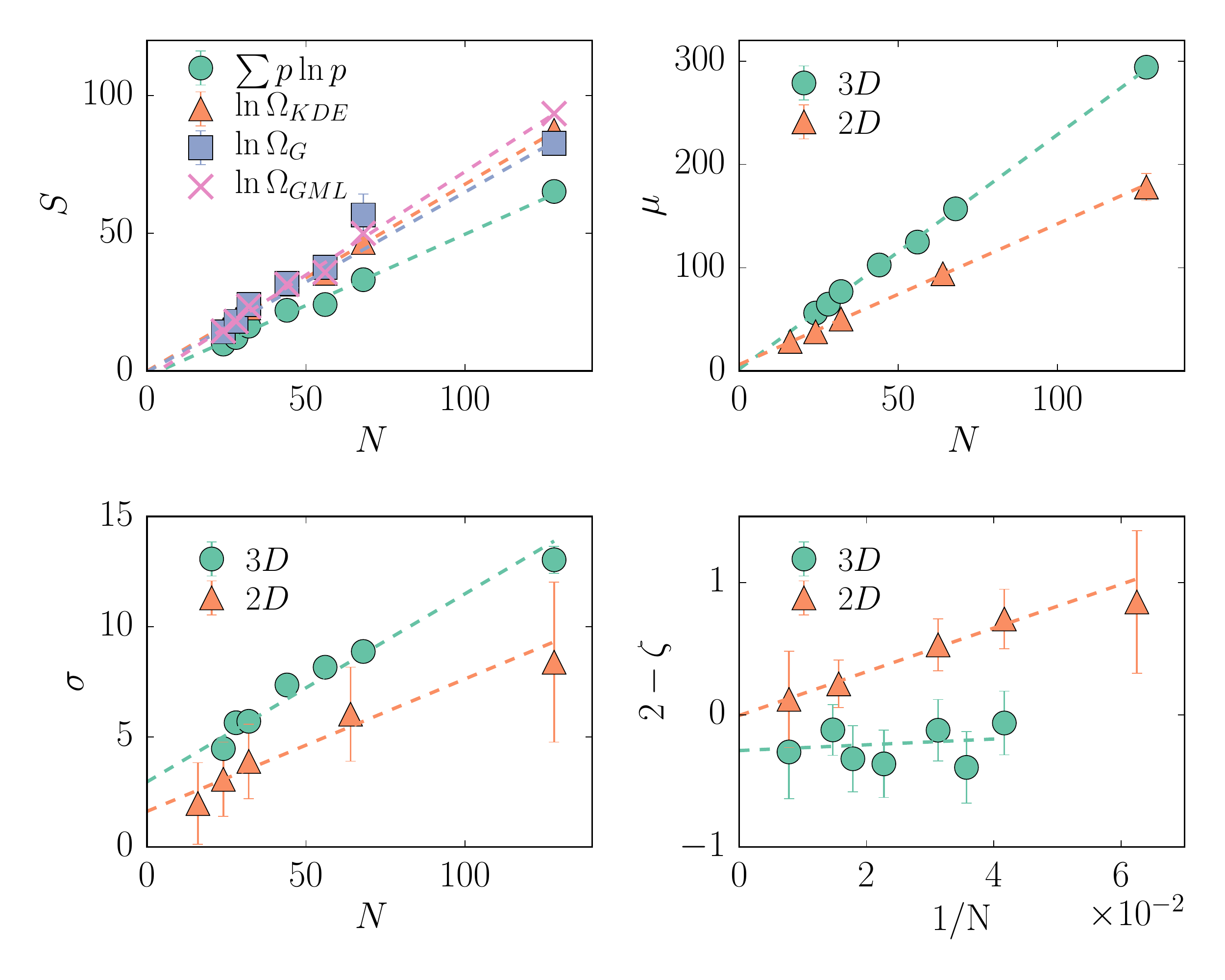}
  \caption{\label{fig::compare_all} Top left, entropy as a function of
    the system size $N$ computed, in order, according to the Gibbs
    configurational entropy and the Edwards configurational entropy
    using a non-parametric fit by kernel density estimation (KDE), a
    parametric fit to a generalised Gaussian c.d.f. using a non-linear
    least squares method and a fit to the corresponding p.d.f. using
    maximum likelihood (ML). Comparison of generalised Gaussian
    best-fit parameters for 2D (see Ref.~\cite{Asenjo14}) and 3D
    sphere packings: scale parameter $\sigma$ (bottom left) and mean
    log-volume $\mu$ (top right) scale linearly with system size $N$;
    distributions are more peaked for 2D packings. In 2D we observe
    much stronger dependence of the shape parameter $\zeta$ (bottom
    right) as a function of system size than in 3D. }
\end{figure}

We first computed two alternative definitions of entropy: the Gibbs
entropy $S_G = -\sum_{i = 1}^{\Omega}p_i\ln(p_i)-\ln(N!)$ and
Edwards (Boltzmann) entropy $S_B= \ln(\Omega)-\ln(N!)$, where $p_i$
is the probability to sample packing $i$ and $\Omega$ is the total
number of mechanically stable states (or minima in the energy
landscape). A detailed discussion of these definitions is outlined in
Sec.~\ref{sec::volume_distributions_data_analysis}. The results of
these calculations are summarised in Fig.~\ref{fig::compare_all}. Our
results strongly suggest that, also in three dimensions, the entropy
thus defined is extensive. Note that extensivity requires not only
that the entropy scales linearly with system size, but also that it
crosses zero at the origin. The slightly higher value of the Edwards
entropy compared to the Gibbs entropy is consistent with the
observation that Edwards' equiprobability corresponds to the maximum
possible entropy of a system with $\Omega$ states.  We also show that
our estimates for the Edwards' entropy are relatively insensitive to
the precise strategy used to compute it. In
Fig.~\ref{fig::compare_all}, we compare three methods: a parametric
fit to a generalised Gaussian cumulative distribution function
(c.d.f.)  using a non-linear least squares method, a fit to the
corresponding probability density function (p.d.f.) using maximum
likelihood, and a non-parametric fit by kernel density estimation,
which makes no a priori assumption about the shape of the
distribution, other than the choice of the kernel function. We note,
once again, that no post-processing is needed to compute the Gibbs
version of the configurational entropy.  Our results are in line with
those reported by Asenjo \emph{et al.} \cite{Asenjo14} for
two-dimensional systems.

The number of mechanically stable states $\Omega$ required by the
Edwards' definition of entropy is obtained subsequently to fitting the
numerically obtained distribution of log-basin volumes (dimensionless
free energies) to a generalised Gaussian distribution and unbiasing it
appropriately, as described in
Sec~\ref{sec::volume_distributions_data_analysis}. We observe that the
best-fit mean and scale parameters of the generalised Gaussian for the
distribution of dimensionless free energies, $\mu$ and $\sigma$ in
Eq.~(\ref{eq:gen_gaussian}) respectively, are also extensive, which
although in line with what was found in two dimensions, is not {\em a
  priori} obvious. Finally we find that the shape parameter, $\zeta$
in Eq.~(\ref{eq:gen_gaussian}), appears to depend only weakly on
system size. The statistics are poor, but the data are compatible with
the assumption that $\zeta\rightarrow 2$ (Gaussian distribution) as
$N\rightarrow\infty$.  In 2D, the same limiting distribution of
$\zeta$, but with a much stronger size dependence, was observed.

\subsection{\label{sec::r3dgen}Entropy in the generalised Edwards ensemble}
We next consider the situation where the configurational entropy is a
function of both the volume $V$ and the stress tensor $\hat{\Sigma}$
of the system.  The number of packings with fixed $V$ and
$\hat{\Sigma}$ is denoted by $\Omega(\hat{\Sigma}, V) $.

In the generalised Edwards ensemble~\cite{Henkes07,Henkes09,Bi15}, we
fix the variables conjugate to $V$ and $\hat{\Sigma}$, {\em viz.}\ the
compactivity $X$ and the inverse angoricity tensor $\hat{\alpha}$. The
generalised `partition function' can then be written as~\cite{Bi15}:
\begin{equation}
  Z_{\text{dyn}} = \sum_{\nu} \omega(\hat{\Sigma}_{\nu}, V_{\nu})
  e^{-V_{\nu}/X} e^{-\text{Tr}(\hat{\alpha} \hat{\Sigma}_{\nu})},
\end{equation}
where $V_{\nu}$ and $\hat{\Sigma}_{\nu}$ are the volume and the
force-moment (stress) tensor for state $\nu$. The weights $\omega$
account for the protocol dependence of the probability to generate a
state, and the sum runs over all mechanically stable states $\nu$.

We can rewrite this partition function in terms of the density of states: 
\begin{equation}
Z_{\text{dyn}} = \prod_{l,k>l} \iint \dif \hat{\Sigma}^{lk} \dif V
\Omega_{\text{dyn}}(\hat{\Sigma}, V) e^{-V/X}
e^{-\text{Tr}(\hat{\alpha} \hat{\Sigma})}.
\end{equation}
For a system under hydrostatic pressure, and in the absence of shear,
we can write the force-moment tensor as $\hat{\Sigma}=\hat{I} \Gamma$,
where $\Gamma=\mathcal{P}V=\mathrm{Tr}(\hat{\Sigma})/3$ is the
internal Virial of the system. The inverse angoricity tensor
$\hat{\alpha}$ becomes a scalar $\alpha = \partial S / \partial
\Gamma$ \cite{Henkes09}. This result allows to simplify the notation
significantly and at fixed volume, through the mean basin volume
method, we obtain the number of states integrated over all pressure
states, $\Omega(V) = \int \dif{\mathcal{P}} \Omega(V,\mathcal{P})$. We
now discuss how to generalise this procedure so that one can compute
$\Omega(V,\mathcal{P})$, and therefore the configurational entropy, in
the context of the generalised Edwards ensemble.

\subsubsection{Pressure to basin volume power-law relation}
\begin{figure}[t]
    \includegraphics[width=\columnwidth]{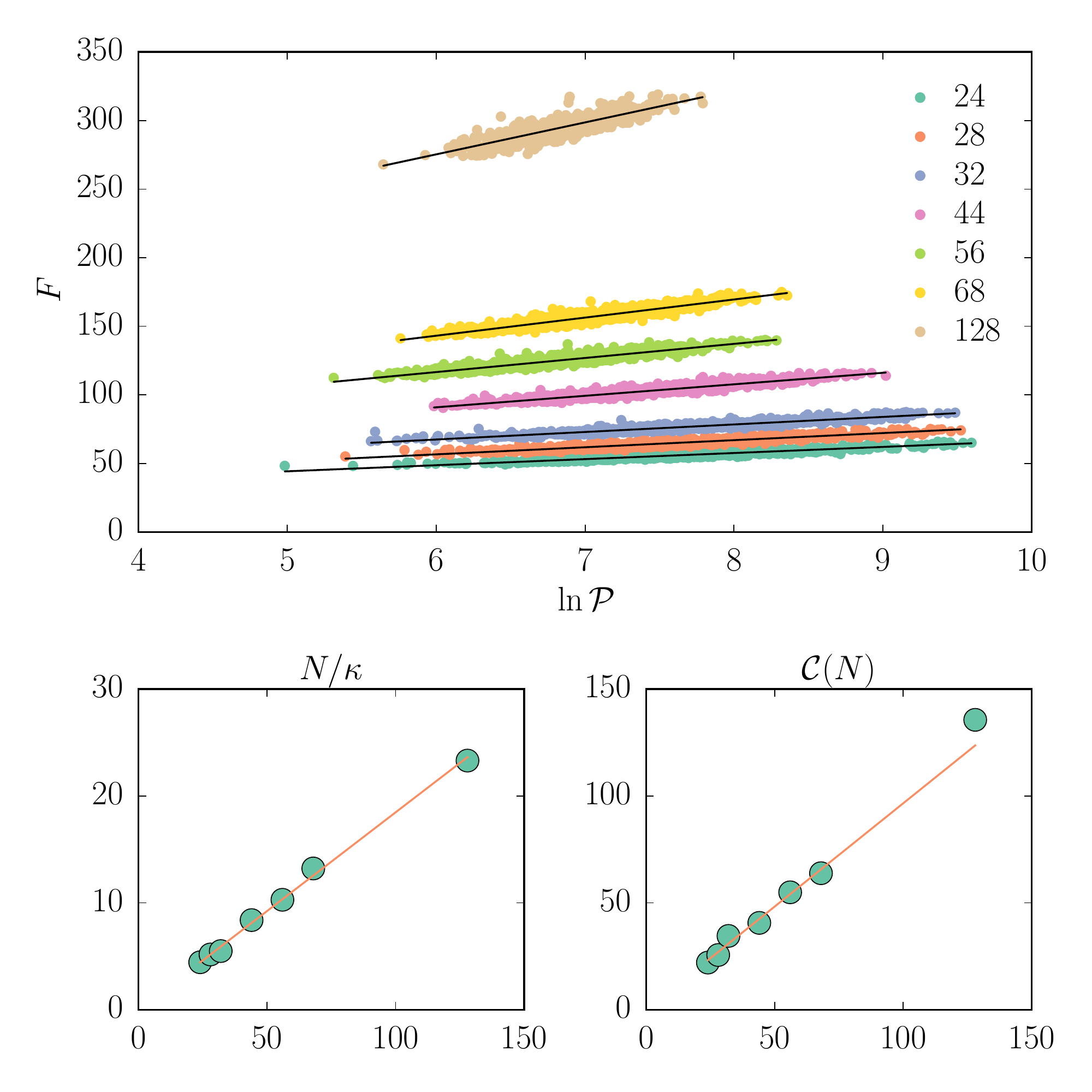}
    \caption{\label{fig::f_logp} Top: dimensionless free energy versus
      pressure of mechanically stable states at fixed volume for
      several system sizes. Best fit lines are in black. In the bottom
      left and right plots we show slope and intercept for each of the
      best fit lines as a function of system size. Both slope and
      intercept scale linearly with system size.}
\end{figure}
To compute $Z(X,\alpha)$ directly, we would have to evaluate
$\Omega(\mathcal{P}, V) $ as a function of both $\mathcal{P}$ and
$V$. Whilst, with the tools that we have, this calculation is in
principle possible, the computational costs would be several orders of
magnitude larger than the, already quite substantial, costs of
computing $\Omega(V)$. This would suggest that the computation of
$Z(X,\alpha)$ is not possible at present.

However, it turns out that we can still estimate the generalised
configurational entropy because, as we discuss below, we observe a
surprisingly strong correlation between pressure and basin volume.

From Fig.~\ref{fig::f_logp} we see that the basin volume for a given
pressure state at fixed volume is strongly correlated with the
pressure $\mathcal{P}$.  As the figure suggests, the relation between
$-\ln (v)$ and $\ln (\mathcal{P})$ is approximately linear, and
hence
\begin{equation}
\label{eq:pressures_f_law}
F(\mathcal{P} | N, \phi_\text{SS}) \equiv -\ln(v) =
\frac{N}{\kappa}\ln(\mathcal{P}) + \mathcal{C}(N),
\end{equation}
where $\kappa$ denotes the slope of the linear fit, and
$\mathcal{C}(N)$ its value at ${\mathcal P}=1$ (see
Fig.~\ref{fig::f_logp}). The value of $\kappa$ is not known a
priori. It seems likely that $\kappa$ depends on the functional form
of the potential. $\mathcal{C}(N)$ is a even less universal linear
function of $N$, as it depends on the choice of units.  

We anticipate that this power law relationship survives for packings
in two dimensions for a wide spectrum of packing fractions $\phi >
\phi_J$ \emph{viz.} as long as the system is jammed and sufficiently
over-compressed \footnote{Results to appear elsewhere.}.

\subsubsection{Gibbs configurational entropy}
\begin{figure}[t]
  \includegraphics[width=\columnwidth]{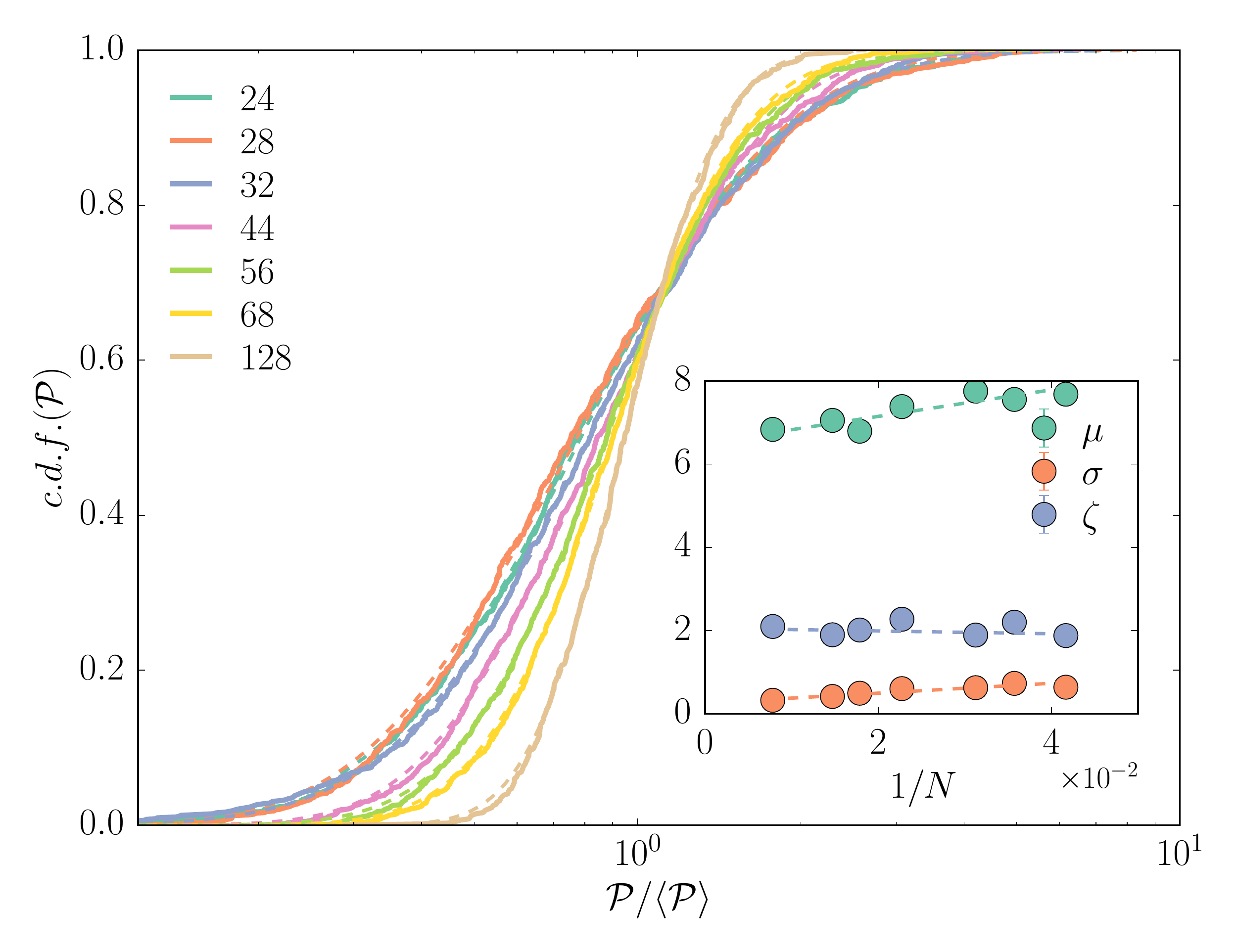}
  \caption{\label{fig::gen_fits} Empirical cumulative distribution
    functions of the pressures for several system sizes. Dashed lines
    in the corresponding colour are curves of best fit to a
    generalised log-normal distribution. The curves are mostly
    indistinguishable. Inset: best fit parameters for the generalised
    log-normal distribution as a function of system size. The mean
    $\mu$ and scale parameter $\sigma$ scale linearly with $1/N$,
    while the shape parameter $\zeta$ is approximately insensitive
    with respect to system size.}
\end{figure}
\begin{figure}
    \includegraphics[width=\columnwidth]{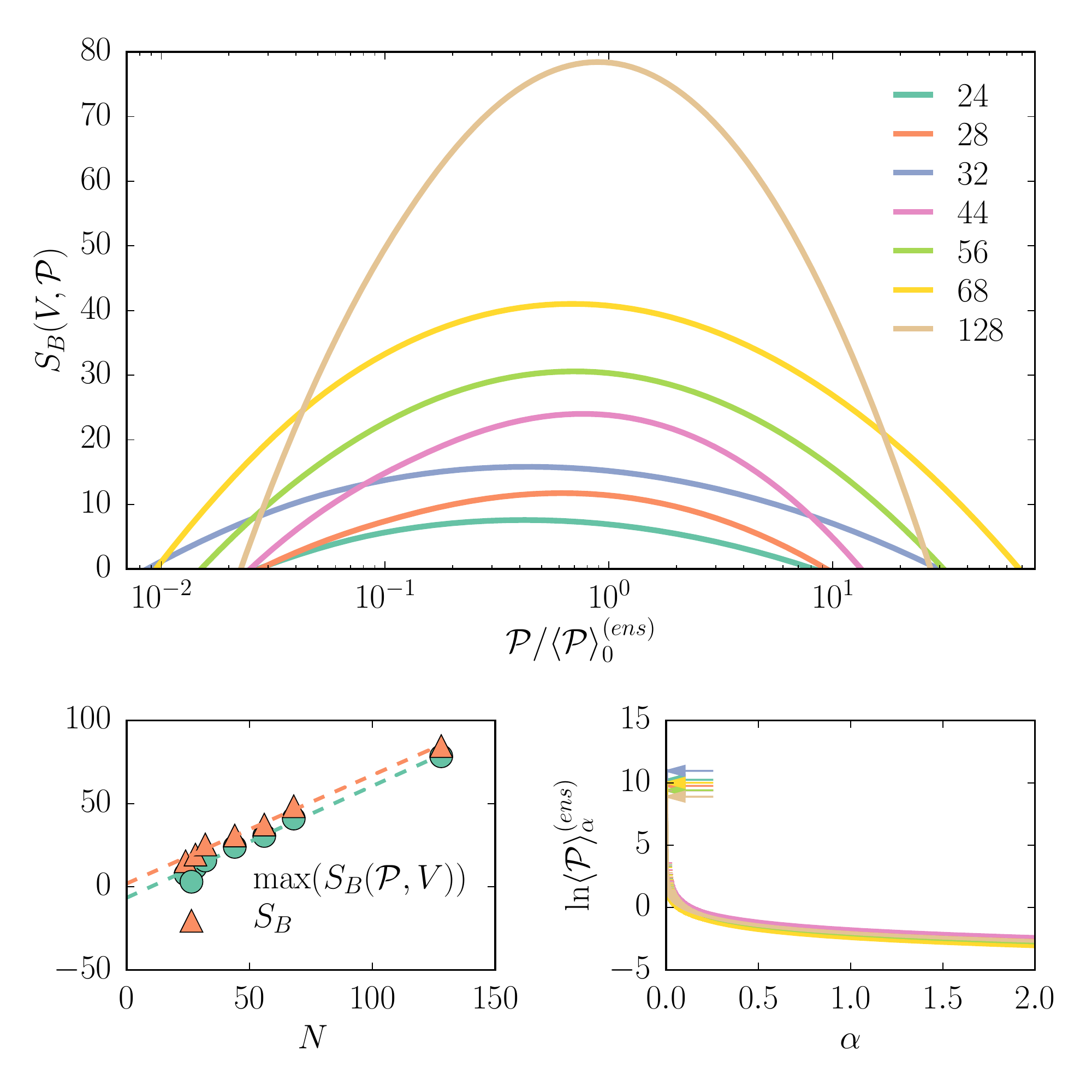}
    \caption{\label{fig::p_S} Top: generalised Edwards entropy at
      fixed volume fraction for various system sizes. The curves show
      a well defined maximum for all sizes, while their shape depends
      on the specific parameters of the generalised log-normal that
      best fits the underlying distribution of pressures. Bottom left:
      comparison between the Edwards entropy and the maximum value
      attained by each curve: $\max[S_B(\mathcal{P},V)]$ scales
      linearly with size and its value is progressively closer to the
      marginal (total) Edwards entropy $S_B(V)$, consistent with the
      fact that $S_B(\mathcal{P},V)$ is a negative exponential
      function, and the area under the curve is dominated by the mode
      for increasing system size. $S_B(V)$ should constitute an upper
      limit to $\max[S_B(\mathcal{P},V)] \leq S_B(V)$ and the two
      should be equivalent only in the thermodynamic limit. Bottom
      right: ensemble average of the pressure computed as a function
      of inverse angoricity $\alpha$ and system size. The curves, in
      the same colour as the top figure, do not diverge and the arrows
      indicates their value at $\alpha=0$.}
\end{figure}
Using our approximate relation between pressure and basin volume, we
can now rewrite Eq.~(\ref{eq:pressures_f_law}) in terms of the
probabilities for each jammed state
\begin{equation}
\label{eq:pressures_f_law_prob}
\ln(p_i) = -\frac{N}{\kappa}\ln(\mathcal{P}_i) - \mathcal{C}(N) -
\ln(\mathcal{V}),
\end{equation}
which when substituted in the definition for the Gibbs entropy
Eq.~(\ref{eq::apf_shannon}), gives the configurational entropy at a
given volume in terms of the biased mean log-pressure
\begin{equation}
    S_G = \frac{N}{\kappa} \left\langle \ln(\mathcal{P})
    \right\rangle_{\mathcal{B}} + \mathcal{C}(N) + \ln(\mathcal{V}) -
    \ln(N!).
\end{equation}
The significance of this equation should be apparent: for a
sufficiently over-compressed packings of soft spheres at a given
packing fraction, the Gibbs configurational entropy can be
approximately computed from sole knowledge of the average pressure,
provided that $\kappa$ is known.

\subsubsection{Generalised Edwards configurational entropy}

To recover the number of states as a function of volume and stress we
note that
\begin{equation}
  \label{eq:omega_volume_stress}
  \Omega(V, \mathcal{P}) = \Omega(V)
  \int_{\mathcal{P}}^{\mathcal{P}+\delta\mathcal{P}}\mathcal{U}(x|V)\dif
  x,
\end{equation}
where $\mathcal{U}(\mathcal{P}|V)$ is the unbiased probability
distribution function of stresses at some specified volume. The
directly measured distribution of pressures depends on the protocol
with which packings are generated.

We distinguish between the biased,
$\mathcal{B}(\mathcal{P}|N,\phi_\text{SS})$ (as sampled by the packing
protocol), and the unbiased,
$\mathcal{U}(\mathcal{P}|N,\phi_\text{SS})$, pressure distributions.
Since the configurations were sampled proportional to the volume of
their basin of attraction, using Eq.~(\ref{eq:pressures_f_law}) we can
compute the unbiased distribution analogously to
Eq.~(\ref{eq:unbiasing}) as
\begin{equation}
  \label{eq:unbiased_p_dist}
    \mathcal{U}(\mathcal{P}|N,\phi_\text{SS}) = \mathcal{Q}(N,\phi_\text{SS})
    \mathcal{B}(\mathcal{P}|N,\phi_\text{SS}) e^{\mathcal{C}(N)}
    \mathcal{P}^{N /\kappa},
\end{equation}
where $\mathcal{Q}(N,\phi_\text{SS}) = \langle v
\rangle(N,\phi_\text{SS})$ is the normalisation constant.

Upon substitution of $\ln\Omega(V) = \ln(\mathcal{V}) - \ln(\langle
v \rangle)$ and of Eq.~(\ref{eq:unbiased_p_dist}) for
$\mathcal{U}(x|V)$, we write an expression for the Edwards entropy as
a function of volume and pressure
\begin{equation}
\begin{aligned}
  \label{eq:gen_edw_entropy}
  S_B(V, \mathcal{P}) = & \ln \left(
  \int_{\mathcal{P}}^{\mathcal{P}+\delta\mathcal{P}}\mathcal{B}(x|V)x^{N/\kappa}\dif
  x\right) \\& + \ln(\mathcal{V}) + \mathcal{C}(N) - \ln(N!).
\end{aligned}
\end{equation}
We fit the empirical cumulative distribution function (c.d.f.)~of
$\mathcal{B}(\mathcal{P})$ with the generalised log-normal
c.d.f. corresponding to Eq.~(\ref{eq:gen_lognormal}) (see
Fig.~\ref{fig::gen_fits}). We then numerically evaluate the
generalised Edwards entropy $S_B(\mathcal{P},V)$ at fixed volume, as
shown in Fig.~\ref{fig::p_S}.

In the thermodynamic limit we find
\begin{equation}
\label{eq:entropy_per_particle2}
s_B(\phi_\text{SS}) = 1 + c + \frac{\langle \ln (\mathcal{P})
  \rangle_{\mathcal{B}}}{\kappa} - \ln(\phi_\text{SS}) -
f_\text{ex}(\phi_\text{HS}),
\end{equation}
where $c=\mathcal{C}(N)/N$, see Appendix \ref{sec::gen_edwards} for
details of the derivation and further discussion.

In Fig.~\ref{fig::p_S} we also show the predicted expectation value
for the pressure obtained via the ensemble average at arbitrary
inverse angoricity $\alpha$,
\begin{equation}
\label{eq:ens_avg}
\langle \mathcal{P} \rangle_{\alpha}^{(ens)} =
\frac{\displaystyle\int_0^{\infty}\mathcal{P}
  \mathcal{B}(\mathcal{P}|V)\mathcal{P}^{N/\kappa}e^{-\alpha\mathcal{P}V}\dif
  \mathcal{P}}{\displaystyle\int_0^{\infty}\mathcal{B}(\mathcal{P}|V)\mathcal{P}^{N/\kappa}e^{-\alpha\mathcal{P}V}
  \dif \mathcal{P}}.
\end{equation}
 
\makeatletter{}\section{\label{sec::packing_sampling_protocol}Packing preparation
  protocol}
\subsection{\label{sec::sampling_packings}Sampling packings}

The physical properties of granular packings may depend strongly on
the preparation protocol.  This is illustrated by the
Lubachevsky-Stillinger algorithm (LSA) procedure to prepare jammed
packings of hard particles~\cite{Lubachevsky91} by compression (or,
equivalently, by `inflation' of the particles).  If a monodisperse HS
fluid is compressed rapidly the LSA will generate a low
volume-fraction disordered packing.  However, for (very) slow
compression rates, LSA will produce dense crystals
\cite{Lubachevsky91, Torquato00}.

In the present work, we study a fluid of polydisperse spheres. We use
a protocol related to a Stillinger-Weber quench that maps each fluid
state to a local minimum, or ``inherent structure'', connected by a
path of steepest descent \cite{Stillinger82, Stillinger84}.

To prepare the polydisperse fluid, we draw $N$ particle radii
$\{r\}_N$ from a Gaussian distribution
$\mathrm{Normal}(1,\sigma_\text{HS}) > 0$, truncated at $r=0$ (note
that in our application the standard deviation $\sigma_\text{HS}$ is
sufficiently small that it is extremely improbable to ever sample a
negative radius).  We set the box size to meet the target packing
fraction of the hard sphere fluid $\phi_\text{HS}$ and then place the
particles in a random valid initial hard spheres configuration. The
initial configuration is then evolved by a MC simulation \cite{mcpele}
consisting of single particle random displacements and
particle-particle swaps, and after equilibration, new configurations
are recorded at regular intervals. We choose the length of these
intervals such that, on average, each particle diffuses over a
distance equal to the diameter of the largest particle.  As long as
$\phi_\text{HS}$ is well below the volume fraction where the fluid
undergoes structural arrest, the allowed configurations of the fluid
can be sampled uniformly.  Importantly, this volume fraction is well
below the random close packing ($\phi_\text{HS}^{(\text{RCP, 3D})}
\approx 0.64$ and $\phi_\text{HS}^{(\text{RCP, 2D})} \approx 0.82$
\cite{OHern03}).

Given these HS fluid configurations, we now switch on the soft,
repulsive interaction to generate over-compressed jammed packings of
the particles (see Fig.~\ref{fig:3d_packings}).  The particles are
inflated with a WCA-like potential \cite{Weeks71} to reach the target
soft packing fraction $\phi_\text{SS} > \phi_\text{HS}^{(\text{RCP})}
> \phi_\text{HS}$.  The hard spheres are inflated proportional to
their radius, so that the soft sphere radius is
\begin{equation}
    r_s
    =\left(\frac{\phi_\text{SS}}{\phi_\text{HS}}\right)^{1/d}
    r_h,
\end{equation}
where $d$ is the dimensionality of the box. Clearly, this procedure
does not change the polydispersity of the sample.
\subsection{\label{sec:wca_pot}Soft shells and minimisation}
We define the WCA-like potential around a hard core as follows:
consider two spherical particles with hard core distance $r_h$ and
soft core contact distance $r_s=r_h(1+\theta)$, with $\theta =
(\phi_\text{SS}/\phi_\text{HS})^{1/d}-1$. We can then write a
horizontally shifted hard-sphere plus WCA (HS-WCA) potential as
\begin{equation}
  \label{eq:hswca}
    v_\text{HS-WCA}(r) = 
    \left\{ 
    \begin{array}{l l}
        \infty &\quad r \leq r_h,\\
        \begin{array}{l}
        4\epsilon \left[
          \left(\frac{\displaystyle\sigma(r_h)}{\displaystyle
            r^2-r_h^2} \right)^{12}
          \right. \\ \left. -\left(\frac{\displaystyle\sigma(r_h)}{\displaystyle
            r^2-r_h^2} \right)^6 \right] + \epsilon
        \end{array}
        &\quad r_h < r < r_{s}, \\
        0 &\quad r \geq r_s
        \end{array}
    \right.
\end{equation}
where $\sigma(r_h) = (2\theta +\theta^2) r_h^2/2^{1/6}$ guarantees
that the potential goes to zero at $r_s$. For computational
convenience (avoidance of square-root evaluations), the potential in
Eq.~\ref{eq:hswca} differs from the WCA form in that the
inter-particle distance in the denominator of the WCA potential has
been replaced with a difference of squares. Note that this implies
that our potential resembles a 6-3 potential more than a 12-6
potential. For our purpose, this difference is immaterial: we just
need a short-ranged repulsive potential that diverges at the hard-core
diameter and vanishes continuously at the soft-core diameter.  The
functional form of this potential is very similar to the HS-WCA
potential used by Asenjo \emph{et al.}  \cite{Asenjo14}, but cheaper
to compute.  We note that this potential is a $C^{1}$ type function,
that is, its first derivative is continuous but not differentiable and
its second derivative is discontinuous at $r_s$. We take advantage of
this property for the identification of rattlers (non-jammed
particles) in our packings.

Numerically evaluating this potential, we match the gradient and
linearly continue the function $v_\text{HS-WCA}(r)$ for $r \leq r_h +
\varepsilon$, with $\varepsilon > 0$ an arbitrary small constant, such
that minimisation is still meaningful if overlaps do occur.

The HS-WCA pair-potential was implemented using cell-lists
\cite{allen1989computer, pele} with periodic boundary conditions,
guaranteeing $\mathcal{O}(N)$ time complexity to the energy and
gradient evaluations. Energy minimisations were performed with the
CG$\_$DESCENT algorithm \cite{Hager05, Hager06, PyCG_DESCENT} which,
compared to FIRE \cite{Bitzek06, pele}, reduces the average number of
function evaluations for our system by a factor of $5$, while
preserving many of its desirable properties.

\makeatletter{}\section{\label{sec::basin_volume_thermodynamic_integration}Basin
  volume by thermodynamic integration}
The basin of attraction of a given minimum-energy configuration is the
collection of all points connected to that minimum via a path of
steepest descent \cite{Mezey82, Wales03}. To measure the volume of a
basin of attraction in the PES, we use thermodynamic integration
\cite{Frenkel84, Polson00} and parallel tempering (PT)
\cite{Lyubartsev92, Marinari92, Geyer95, mcpele}.

The basic idea behind the method is that, but for the sign, the
logarithm of the basin volume can be viewed as a dimensionless free
energy.  We cannot determine this free energy directly. We now switch
on an increasingly harmonic potential that has its minimum at the
minimum of the basin. In the limit of very large coupling constants
(how large depends on the shape of the basin) the boundaries of the
basin no longer affect the free energy of the system, which has
effectively been reduced to a $dN$ dimensional harmonic oscillator
with known free energy (for more details, see
Appendix~\ref{sec::Einstein_crystal_CM}). For zero coupling constant,
instead, the system is completely unconstrained and therefore in the
state of interest. Thermodynamic integration allows us to compute the
free energy difference between a reference state of known free energy
and the (unknown) free energy associated with the original basin of
attraction.

A closely related approach is often used to compute the free energy of
crystals of particles with a discontinuous potential, such as hard
spheres~\cite{Frenkel84, Polson00, Frenkel02}.  Details of that method
are summarised in the Appendix~\ref{sec::flm}, and the extension of
the technique to basin volume measurement is described below. Details
of the Hamiltonian PT are discussed in
Appendix~\ref{sec::sampling_th_integrand}.
\subsection{\label{sec::flm_for_basins}Free energy calculation for
  basin volumes}

To measure the volume of a basin by thermodynamic integration, we
perform a walk inside the basin, that is, we start the MCMC random
walk from the minimum energy configuration $\mathbf{r}_i$ and we
reject every move that takes us outside the basin \cite{Xu11,
  Frenkel13, Asenjo14}.  This procedure can be cast in normal Monte
Carlo language by defining an effective potential energy function
(oracle) $U_{B}(\mathbf{r} | \mathbf{r}_i)$ which is zero inside the
basin and infinite outside.  We can then write the volume of the
basin:
\begin{equation} 
    v_i = \int \dif \mathbf{r} e^{-U_{B}(\mathbf{r}| \mathbf{r}_i)}.
\end{equation}
In order for the oracle to test whether a proposed configuration is
inside or outside the basin, a full energy minimisation must be
performed. The numerous potential energy calls required for a full
energy minimisation represent the major obstacle to the scalability of
the method.

We view the negative log-basin-volume as a dimensionless free energy
$F_i \equiv -\ln(v_i)$~\cite{Xu11} and compute it by thermodynamic
integration, as described in Appendix \ref{sec::flm}.  Therefore, we
write, analogously to Eq.~(\ref{eq:int_frenkelladd}):
\begin{equation}
    \label{eq:bv_1}
    - \ln v_i = F_{\mathrm{har}}(k_\text{max}) - \frac{1}{2}
    \int_0^{k_\text{max}} \dif k \left \langle \vert \mathbf{r} -
    \mathbf{r}_i \vert \right \rangle_k ,
\end{equation}
where $\mathbf{r}_i$ denotes the coordinates of the $i$-th energy
minimum.  Unless $k_\text{max}$, the maximum spring constant of the
harmonic reference system, is very large, a finite fraction of the
points belonging to the purely harmonic reference system will be
located in the region where $U_B=\infty$.

We can correct for this effect in our calculation of
$F_{\mathrm{har}}(k_\text{max})$ by computing the ratio of the
partition functions of a system with a harmonic spring constant
$k_\text{max}$, both with and without the basin potential energy
function $U_B$.  This ratio is given by
\begin{equation}
    \label{eqn::rho_kmax}
\mathcal{R} \equiv \frac {\displaystyle\int \dif \mathbf{r} \exp[-V(\mathbf{r}
        | \mathbf{r}_i, k_\text{max}) - U_B(\mathbf{r} |
        \mathbf{r}_i)]} {\displaystyle\int \dif \mathbf{r}
      \exp[-V(\mathbf{r} | \mathbf{r}_i, k_\text{max})]},
\end{equation}
where $V$ is the sum of the hard-core potential and the harmonic
potential with spring constant $k_\text{max}$, see
Eq.~(\ref{eq::hs_plus_harmonic}).  We note that $\mathcal{R}$ can be
computed using a `static' (i.e. non-Markov chain) Monte Carlo
simulation, sampling directly from the Boltzmann distribution of the
harmonic oscillator with spring constant $k_\text{max}$.  Since the
integral in the denominator is known [see Eq.~(\ref{eq:z_har})], we
write the dimensionless free energy of the harmonic reference state
for basin $i$ as
\begin{equation}
    \label{eq:f_har_cor}
    F_{\text{har}}(k_\text{max}) = -\frac{dN}{2}\ln\left(
    \frac{2\pi}{k_\text{max}}\right) - \ln \mathcal{R}.
\end{equation}

We note that, in order to avoid a singularity in the integrand, it is
useful to perform the simulations fixing the centre of mass.  It
follows that the same corrections to the free energy as derived in
Refs.\cite{Frenkel84, Polson00, Frenkel02} must be applied: similarly
to Eq.~(\ref{eq:full_f2}), but with the additional correction in
Eq.~(\ref{eq:f_har_cor}), we write the basin volume as:
\begin{equation}
    \label{eq:bv_cm}
    \begin{aligned}
    -\ln v_i &= \Delta F^{(\text{CM})} -\ln
    \left(V_\text{box}\right) \\ &- \frac{(N-1)d}{2}\ln \left(
    \frac{2\pi}{k_\text{max}}\right) -\ln\mathcal{R},
    \end{aligned}
\end{equation}
where $\Delta F^{(\text{CM})}$ is the integral in Eq.~(\ref{eq:bv_1}),
and the ensemble averages have been computed with a constrained centre
of mass and it is evaluated as in Eq.~(\ref{eq:int_cov3}).

\begin{figure}[t]
    \includegraphics[width=\columnwidth]{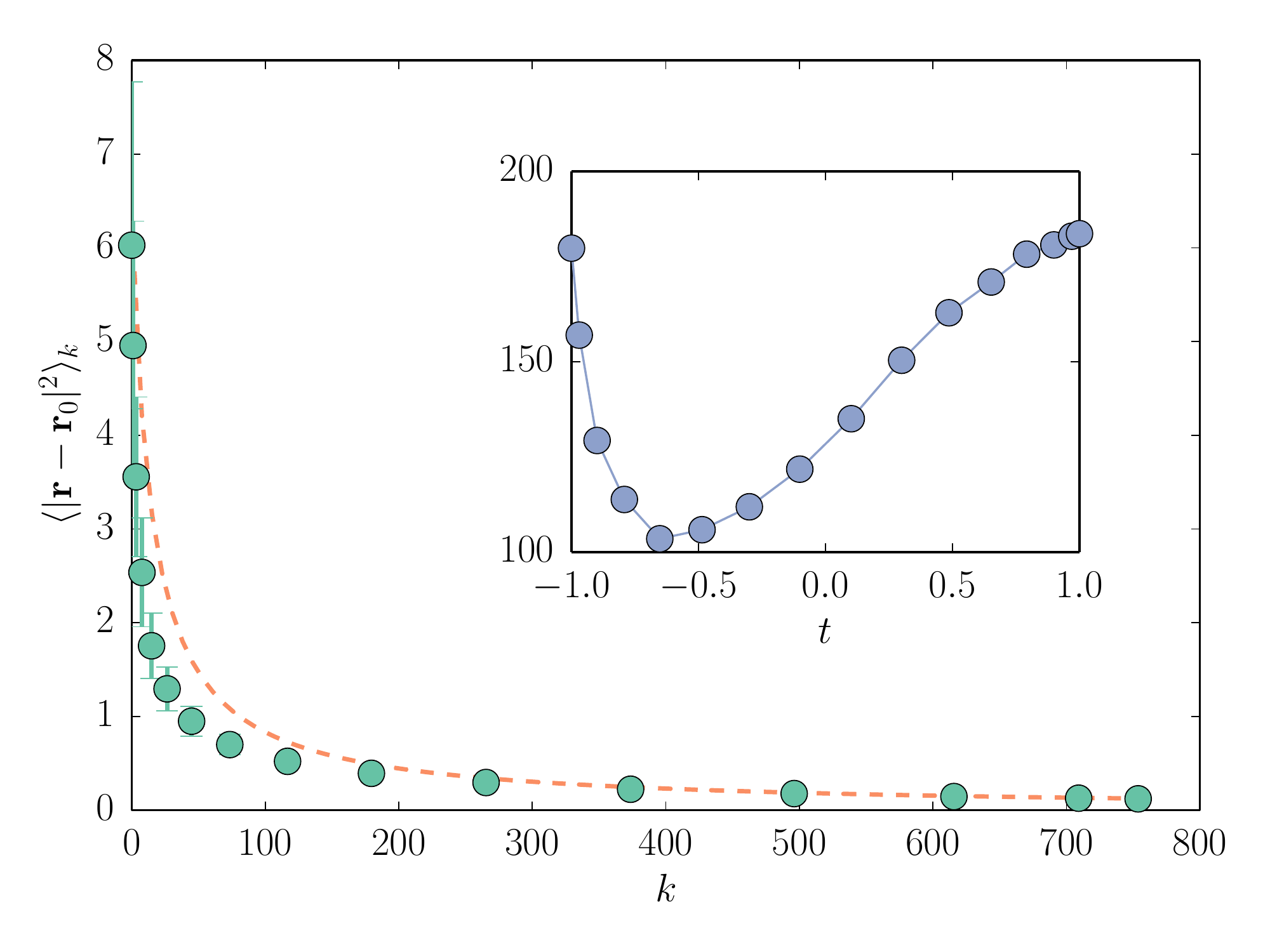}
    \caption{\label{fig::u2_vs_k} Average squared displacement $\left
      \langle |\mathbf{r}-\mathbf{r}_0|^2 \right \rangle_{k}$ as a
      function of the spring constant $k$ (symbols).  The dashed line
      shows the expression in Eq.~(\ref{eq:rms_approx_k}).  The data
      is measured for a packing of $N=32$ spheres, with
      $\phi_\text{HS}=0.5$ and $\phi_\text{SS}=0.7$ via Hamiltonian
      PT. Inset: corresponding integrand for the thermodynamic
      integration, resulting from the change of variables in
      Eq.~(\ref{eq:int_cov3}).}
\end{figure}

Figure~\ref{fig::u2_vs_k} shows an example of the mean squared
displacement $\left \langle |\mathbf{r}-\mathbf{r}_0|^2 \right
\rangle_{k}$, as a function of the spring constant $k$, along with the
approximate expression in Eq.~(\ref{eq:rms_approx_k}) used to
construct the change of variables in Eq.~(\ref{eq:int_cov3}).  The
resulting integrand, after the variable transform, is shown in the
inset of Fig.~\ref{fig::u2_vs_k}.

\makeatletter{}\section{\label{sec::volume_distributions_data_analysis}Basin volume
  distributions and data analysis}
Once the volumes of multiple basins have been sampled, these data can
be used to compute the number of distinct packings \cite{Asenjo14},
and from that, the Edwards entropy~\cite{Edwards89}.  Furthermore we
analyse the distribution of pressures of the different energy minima
at given volume. In this work, we express pressure and volume in
reduced units $\mathcal{P}/\mathcal{P}^{\ast}$ and $v/v^{\ast}$
everywhere with $v^{\ast} \equiv (4\pi/3)\langle r_h ^3\rangle$ and
$\mathcal{P}^{\ast}\equiv \epsilon/v^{\ast}$ being the units of volume
and pressure, respectively.
\subsection{Gibbs configurational entropy}
Let us first consider the `Gibbs' configurational entropy, $S_G$, defined by
Asenjo \emph{et al.} \cite{Asenjo14}:
\begin{equation}
    \label{eq::basic_shannon}
    S_G = -\sum_{i = 1}^{\Omega} p_i \ln(p_i) - \ln(N!),
\end{equation}
where $p_i$ is the probability to sample packing $i$. For our
preparation protocol, packings are sampled according to the volume of
their basin of attraction, such that $p_i = v_i / \mathcal{V}$.  Then
Eq.~(\ref{eq::basic_shannon}) gives
\begin{equation}
  \begin{aligned}
    \label{eq::apf_shannon}
    S_G &= -\sum_{i = 1}^{\Omega} p_i\ln(v_i) + \ln(\mathcal{V}) -
    \ln(N!) \\
    &= \langle F \rangle_{\mathcal{B}} + \ln(\mathcal{V}) -
    \ln(N!).
    \end{aligned}
\end{equation}
The sum in Eq.~(\ref{eq::apf_shannon}) is the mean of the negative
log-basin volumes (dimensionless free energies), as computed above,
and weighted by the probabilities of preparing the corresponding
basins.  Therefore, the entropy can be obtained directly, and without
approximation, from the sampled mean basin dimensionless free energy.

From Eq.~(\ref{eq::apf_shannon}) we can also write the entropy per
particle in the thermodynamic limit as
\begin{equation}
\label{eq:entropy_per_particle1}
s_G(\phi_\text{SS}) = 1 + \langle f \rangle_{\mathcal{B}} +
\ln(\phi_\text{SS}) - f_\text{ex}(\phi_\text{HS}) ,
\end{equation}
where $f_\text{ex}(\phi_\text{HS})$ is the excess free energy of the
hard spheres fluid. In deriving this results we used Stirling's
approximation for large $N$ and the fact that
$V_{\text{box}}/v^{\ast}=N / \phi_\text{SS}$.
\subsection{Edwards configurational entropy}
Edwards~\cite{Edwards89} suggested a Boltzmann-like entropy, where $S$
equals the logarithm of\enskip $\Omega$, the total number of packings.
Asenjo \emph{et al.}~\cite{Asenjo14} showed that, even for
polydisperse particles, indistinguishability of macrostates requires
that
\begin{equation}
\label{eq:boltzmann_entropy}
S_B = \ln(\Omega) - \ln(N!),
\end{equation}
The subtraction of $\ln(N!)$ is necessary to guarantee extensivity of
the entropy. Unlike the Gibbs definition of entropy,
Eq.~(\ref{eq:boltzmann_entropy}) makes the explicit assumption of
equiprobability of states.

For a direct computation of the number of packings $\Omega$, using
Eq.~(\ref{eq:bv_omega}), we need the average basin volume $\langle v
\rangle$.  Since our preparation protocol samples each minimum with a
probability proportional to the volume of its basin of attraction, our
samples of $v$ are biased accordingly.  Therefore, to obtain the
unbiased average basin volume $\langle v \rangle$, the sampled basin
volume distribution needs to be unbiased \cite{Frenkel13, Asenjo14,
  Paillusson15}.  The unbiasing method used in the following work
requires an analytical (or at least numerically integrable)
description of the biased basin free energy distribution function.
Different approaches to modelling this distribution give rise to
somewhat different analysis methods, which all yield consistent
results. Again, we stress that no such additional steps are needed to
compute the `Gibbs' version of the configurational entropy.

We distinguish between the biased, $\mathcal{B}(F|N,\phi_\text{SS})$
(as sampled by the packing protocol), and the unbiased,
$\mathcal{U}(F|N,\phi_\text{SS})$, free energy distributions.  Since
the configurations were sampled proportional to the volume of their
basin of attraction, we can compute the unbiased distribution as
\begin{equation}
    \label{eq:unbiasing}
    \mathcal{U}(F|N,\phi_\text{SS}) = \mathcal{Q}(N,\phi_\text{SS})
    \mathcal{B}(F|N,\phi_\text{SS}) e^F
\end{equation}
where $\mathcal{Q}(N,\phi_\text{SS})$ is the normalisation constant
\begin{equation}
    \label{eq::un_bias_integral_c_omega}
    \mathcal{Q}(N, \phi_\text{SS}) = \left [
      \int_{F_\text{min}}^{\infty} \dif F
      \mathcal{B}(F|N,\phi_\text{SS})e^F \right]^{-1} = \langle v
    \rangle(N,\phi_\text{SS}).
\end{equation}
From Eq.~(\ref{eq:unbiasing}), unbiasing of the raw free energy
distribution seems straightforward, however Asenjo \emph{at al.}
\cite{Asenjo14} noted that the most probable basins are about
$O(10^3)$ more probable than the small ones.  Upon unbiasing, this
factor is multiplied by a factor of about $e^{-20}$, hence they
observe that small basins are much more numerous than large ones and
grossly under-sampled.

To overcome this problem, one can fit the biased measured free energy
distribution $\mathcal{B}(F|N,\phi_\text{SS})$ and perform the
unbiasing via Eq.~(\ref{eq::un_bias_integral_c_omega}) on the best
fitting distribution.  $\mathcal{B}(F|N,\phi_\text{SS})$ must be
bounded, hence it should decay with a functional form $\exp(-F^{\nu})$
where $\nu > 1$.

Before performing the fit we remove outliers from the free energy
distribution following the distance-based outlier removal method
introduced by Knorr and Ng \cite{Knorr98}. This is a form of
clustering for which we choose to keep only those points for which at
least half of the remaining data set is within $3\sigma$ from the
point, where $\sigma$ is the standard deviation computed for the raw
data set. This procedure typically results in the exclusion of one or
two points and it is essential for a successful fit to a generalised
Gaussian model.
\subsubsection{Generalised Gaussian}
Assuming that $\mathcal{U}(F|N,\phi_\text{SS})$ is unimodal, which has
been verified for very small systems \cite{Xu11}, one can fit the raw
distribution $\mathcal{B}(F|N,\phi_\text{SS})$ with a three-parameter
generalised normal distribution
\begin{equation}
    \label{eq:gen_gaussian}
    p(F|\overline{F},\sigma,\zeta) \equiv
    \frac{\zeta}{2\sigma\Gamma(1/\zeta)}\exp\left[-\left(\frac{|F-\overline{F}|}{\sigma}\right)^{\zeta}\right],
\end{equation}
where $\Gamma(x)$ is the gamma function, $\sigma$ is the scale
parameter, $\zeta$ is the shape parameter and $\overline{F}$ is the
mean (free energy) with variance $\sigma^2
\Gamma(3/\zeta)/\Gamma(1/\zeta)$.  In the limit $\zeta \rightarrow 2$
we recover the Gaussian distribution with standard deviation $\sigma$.
In practice it appears to be most stable to fit the empirical biased
cumulative distribution function, rather than the histogram shape
\cite{Asenjo14}.  Alternatively, we also tested fitting to the
observed p.d.f. with the maximum-likelihood method, obtaining
consistent, but more scattered, results (see also
Sec.~\ref{sec::r3d}).
\subsubsection{Kernel density estimate}
To relax the assumption that the empirical distributions can be fitted
by a generalised Gaussian, one can also describe the distributions by
kernel density estimation \cite{Bishop09, scikitKDE}.  Bandwidth
selection is then done using Silverman's rule of thumb as the initial
guess for integrated squared error cross-validation \cite{Bowman84}.
The numerical integration step is performed, as for the generalised
Gaussian description, via Eq.~(\ref{eq::un_bias_integral_c_omega}).
\subsection{Distribution of pressures}
In Sec.~\ref{sec::r3dgen} we have established a link between the
pressure of a packing and the volume of its basin of attraction. In
order to compute the entropy as a function of volume and pressure it
is necessary to unbias the distribution of pressures with respect to
the sampling bias $\exp(-F)$, analogous to the previous section. We
choose to describe the distribution of pressures $\mathcal{P}$ using
the generalised log-normal distribution \cite{Singh12}
\begin{equation}
\label{eq:gen_lognormal}
\begin{aligned}
p(\mathcal{P}|\overline{\ln(\mathcal{P})}, \sigma, \zeta) =&
\frac{\zeta/\mathcal{P}}{2^{(\zeta+1)/\zeta}\sigma\Gamma(1/\zeta)} \\ & \exp\left(-\frac{1}{2}\left|\frac{\ln(\mathcal{P})-\overline{\ln(\mathcal{P})}}{\sigma}\right|^\zeta
\right),
\end{aligned}
\end{equation}
with the first term on the r.h.s. being the normalisation constant and
the remaining notation analogous to that of
Eq.~(\ref{eq:gen_gaussian}). For $\zeta=2$ this distribution reduces
to the log-normal distribution.
 
\makeatletter{}\section{\label{sec::conclusion}Conclusions}

The study of a statistical mechanics of granular materials has been
complicated by the impossibility of directly computing fundamental
thermodynamic quantities. In the present paper we have shown that
configurational entropies of three-dimensional packings can, in fact,
be computed.

We have presented a method for the direct enumeration of the
mechanically stable states of systems consisting of up to $128$
frictionless soft three-dimensional spheres and we have shown that a
definition of extensive entropy is possible, in line with the results
for two dimensional systems reported by Asenjo \emph{et
  al.}~\cite{Asenjo14}, with very minor differences in our
observations. The study of 3D packings is computationally demanding:
the computational time required for each packing ranged between $10$
and $10^4$ cpu hours, depending on system size. The present study
therefore required substantial algorithmic optimisation.

We find that there is an approximately linear relationship between the
logarithm of the pressure of a mechanically stable configuration and
the logarithm of the volume of its basin of attraction.

The unexpected power law relationship between pressure and basin
volume provides a way to extend our approach to the generalised
Edwards ensemble. We can analytically unbias the observed distribution
of pressures and compute the entropy as a function of pressure at a
given volume. Hence we have obtained consistent expressions for the
entropy in the thermodynamic limit. Knowledge of this distribution
enables the first direct computation of angoricity.

Tackling the study of granular materials from the energy landscapes
point of view is rather advantageous, although this does not come
without burdens. This sort of approach is limited to soft frictionless
particles, and we expect it to be reliable only at $\phi>\phi_J$ when
the system is at least slightly over-compressed.  Other theoretical
approaches are useful in more limiting situations, see for instance
the discussion of the stress ensemble in the limit $\phi \rightarrow
\phi_J$ by Henkes and Chakraborty \cite{Henkes07, Henkes09} and the
work on the force network ensemble for systems of almost hard grains
\cite{Van09, Tighe10, Tighe11}.
 
\begin{acknowledgments}
We acknowledge useful discussions with Daniel Asenjo, Carl Goodrich,
Silke Henkes, and Fabien Paillusson.  S.M. acknowledges financial
support by the Gates Cambridge Scholarship.  K.J.S. acknowledges
support by the Swiss National Science Foundation under Grant
No. P2EZP2-152188 and No. P300P2-161078. J.D.S. acknowledges support
by Marie Curie Grant 275544. D.F. and D.J.W. acknowledge support by
EPSRC Programme Grant EP/I001352/1, by EPSRC grant EP/I000844/1 (D.F.)
and ERC Advanced Grant RG59508 (D.J.W.)
\end{acknowledgments}
\appendix
\makeatletter{}\makeatletter{}\section{\label{sec::method_summary}Basin volume method summary}
In summary, to count the number of ways spheres can pack into a given
volume, we use the mean basin volume method outlined in
Sec.~\ref{sec::basic_principle}.  We perform the following simulations
and analysis steps to obtain the required results:
\begin{enumerate}
    \item Obtain a number of different snapshots of an equilibrated
      hard sphere fluid at the desired volume fraction
      $\phi_\text{HS}$, as described in
      Sec.~\ref{sec::sampling_packings}.  This procedure fixes the
      number of measured basin volumes.
    \item Over-compress the sphere configuration by adding a soft
      shell.  This compression yields, upon energy minimisation, a
      jammed packing with soft volume fraction $\phi_\text{SS} >
      \phi_\text{HS}$.
    \item Estimate the maximum spring constant for the PT
      simulations, $k_\text{max}$ in Eq.~(\ref{eq:bv_1}), such that
      $\rho$ in Eq.~(\ref{eqn::rho_kmax}) reaches a value between
      $0.85$ and $0.9$.  This is done by direct sampling and also
      gives the value of the average squared displacement for
      $k_\text{max}$, $\left \langle |\mathbf{r}-\mathbf{r}_0|^2
      \right \rangle_{k_\text{max}}$.
    \item Obtain a preliminary estimate of the average squared
      displacement without harmonic tethering, $\left \langle
      |\mathbf{r}-\mathbf{r}_0|^2 \right \rangle_{0}$, by performing a
      MCMC walk in the basin.  Use this result, with the estimate of
      $k_\text{max}$ from the previous step, to determine the spring
      constants $k$ for the PT simulation, using Eqs.~(\ref{eq:t})
      and (\ref{eqn::gl_integration_general_form}).
    \item Perform a PT simulation to sample $\left \langle
      |\mathbf{r}-\mathbf{r}_0|^2 \right \rangle_{k}$, as described in
      Appendix~\ref{sec::sampling_th_integrand}.
    \item Compute the volume in Eq.~(\ref{eq:bv_cm}) for each basin
      and analyse the distributions for all basins, at fixed volume
      fraction and number of particles, as discussed in
      Sec.~\ref{sec::volume_distributions_data_analysis}.  This makes
      use of the total accessible volume, computed in
      Appendix~\ref{sec::poly_fluid}.
\end{enumerate}
Section \ref{sec::r3d} shows examples of the type of results that can
be obtained.  Evaluation and minimisation of potential energy
functions was performed with the pele \cite{pele} and PyCG\_DESCENT
\cite{PyCG_DESCENT} software packages.  Monte Carlo simulations were
performed with the mcpele package \cite{mcpele}.
 
\section{\label{sec::flm}Free energy calculation for solids}
To compute the free energy of a system with discontinuous potential
energy function (e.g., hard disks or hard spheres), we construct a
reversible path to the corresponding Einstein solid (see e.g.~\cite{Frenkel02}).
The harmonic potential with spring constant $k$ is switched on while
maintaining the hard core interactions intact:
\begin{equation}
    \label{eq::hs_plus_harmonic}
    \begin{aligned}
        V(\mathbf{r}|\mathbf{r}_0, k) &= V_{\text{HS}}(\mathbf{r}) + k
        V_{\text{har}}(\mathbf{r}|\mathbf{r}_0) \\ &=
        V_{\text{HS}}(\mathbf{r}) + \frac{1}{2}k
        |\mathbf{r}-\mathbf{r}_0|^2,
    \end{aligned}
\end{equation}
where $\mathbf{r}_0$ are the equilibrium coordinates of the Einstein
crystal and $V_\text{HS}(\mathbf{r})$ denotes the hard core
interactions.  We can then compute the free energy difference between
the Einstein crystal and the hard core system by evaluating the
integral:
\begin{equation}
\label{eq:int_frenkelladd}
F_\text{HS} = F_{\text{har}}(k_\text{max}) - \int^{k_\text{max}}_{0}
\dif k \left \langle \frac{\partial V(\mathbf{r}|\mathbf{r}_0,
  k)}{\partial k} \right \rangle_k .
\end{equation}
As discussed in Appendix~\ref{sec::Einstein_crystal_CM}, we take the
centre of mass to be fixed to avoid numerical issues in the limit $k
\to 0$.  For a system with fixed centre of mass, we write the free
energy difference between the target and the reference state as
\begin{equation}
    \label{eq:dif_f}
    \Delta F^{(\text{CM})} \equiv F^{(\text{CM})} -
    F^{(\text{CM})}_{\mathrm{har}}.
\end{equation}
From the partition function of the Einstein crystal with fixed centre
of mass, Eq.~(\ref{eq:har_zcm}), and for the unconstrained crystal,
Eq.~(\ref{eq:unc_zcm}), we can rewrite Eq.~(\ref{eq:dif_f}) and
rearrange it for the free energy of the unconstrained crystal:
\begin{equation}
\label{eq:full_f1}
    \begin{aligned}
    F &= \Delta F^{(\text{CM})} + \ln
    \left(\mathcal{P}(\mathbf{r}_\text{CM}=\mathbf{0}) \right) \\ &+
    \frac{d}{2} \ln \left(
    \frac{2\pi\sum_i\mu_i}{k_\text{max}}\right) - \frac{Nd}{2}\ln
    \left( \frac{2\pi}{k_\text{max}}\right),
    \end{aligned}
\end{equation}
where the last term is $F_{\text{har}}$ and the second and third terms
on the right hand side are the CM corrections for the unconstrained
and the constrained solid, respectively.  For a system with unit cell
identical to the simulation box (with periodic boundary conditions),
we have $\mathcal{P}(\mathbf{r}_\text{CM}=\mathbf{0})=1/V_\text{box}$.
Assuming that all particles have unit mass we can rewrite
Eq.~(\ref{eq:full_f1}) as
\begin{equation}
    \label{eq:full_f2}
    F = \Delta F^{(\text{CM})} - \ln \left(V_\text{box}\right) -
    \frac{(N-1)d}{2}\ln \left( \frac{2\pi}{k_\text{max}}\right).
\end{equation}
We are only left with $\Delta F^{(\text{CM})}$, which can be found by
evaluating the integral in Eq.~(\ref{eq:int_frenkelladd}).  In order
to do so, we would like the integrand to be a well behaved function,
possibly flat, permitting Gauss-Lobatto (GL) quadrature
\cite{weisstein_lobatto_2014}.  We transform the integration variable
so that
\begin{equation}
    \label{eq:int_cov}
    \begin{aligned}
    \Delta F^{(\text{CM})} &= \int_0^{k_\text{max}} \frac{\dif
      k}{g(k)} g(k) \frac{1}{2}\left \langle
    |\mathbf{r}-\mathbf{r}_0|^2 \right \rangle_{k}^{(\text{CM})} \\ &=
    \int_{G^{-1}(0)}^{G^{-1}(k_\text{max})} \dif
    \left[G^{-1}(k)\right] g(k) \frac{1}{2} \left \langle
    |\mathbf{r}-\mathbf{r}_0|^2 \right \rangle_{k}^{(\text{CM})},
    \end{aligned}
\end{equation}
where $g(k)$ is some function of $k$ and $G^{-1}(k)$ is the primitive
of the function $1/g(k)$.

To choose an appropriate $g(k)$, we note that in
Eq.~(\ref{eq:free_energy_u2_cm}) for very large $k$ the mean squared
displacement for the solid is
\begin{equation}
    \label{eq:msd_kmax}
    \langle |\mathbf{r}-\mathbf{r}_0|^2 \rangle_{k_\text{max}} =
    \frac{(N-1)d}{k_\text{max}}.
\end{equation}
For $k$ other than $k_\text{max}$, we expect the mean squared
displacement to depend on some effective spring constant. Hence we
write
\begin{equation}
    \label{eq:rms_approx_k}
    \left \langle |\mathbf{r}-\mathbf{r}_0|^2 \right \rangle_{k}
    \approx \frac{(N-1)d}{(k+\xi)},
\end{equation}
such that the mean squared displacement at $k=0$ is
\begin{equation}
    \label{eq:msd_kmin}
    \left \langle |\mathbf{r}-\mathbf{r}_0|^2 \right \rangle_{k=0}
    \approx \frac{(N-1)d}{\xi},
\end{equation}
from which we find $\xi = (N-1) d / \langle
|\mathbf{r}-\mathbf{r}_0|^2 \rangle_{k=0}$ [note that we can self
consistently replace this definition for $\xi$ in
Eq.~(\ref{eq:rms_approx_k}) to obtain an approximation for the mean
squared displacement at arbitrary $k$].  We would like the integrand
$g(k)\langle |\mathbf{r}-\mathbf{r}_0|^2 \rangle_{k}$ in
Eq.~(\ref{eq:int_cov}) to be roughly constant.  Given the
considerations above we choose $g(k) \approx k + \xi.$ One can easily
verify that the integrand is now approximately constant.  We can then
rewrite the integral in Eq.~(\ref{eq:int_cov}) as
\begin{equation}
    \label{eq:int_cov2}
    \begin{aligned}
    \Delta F^{\text{(CM)}} =
    \int_{\ln(\xi)}^{\ln(k_\text{max}+\xi)} \biggl \{ (k+\xi) \frac{1}{2} & \left \langle
    |\mathbf{r}-\mathbf{r}_0|^2 \right \rangle_{k}^\text{(CM)} \\ & \dif \left[
      \ln(k+\xi)\right] \biggl \}.
    \end{aligned}
\end{equation}
Finally, to integrate Eq.~(\ref{eq:int_cov2}) by GL quadrature, we
require a variable, $t$, such that the integral upper and lower bounds
are $[-1,1]$:
\begin{equation}
    \label{eq:t}
    t = \frac{2\ln\left( 1 + k/\xi\right)-1}{\ln\left( 1+
      k_\text{max}/\xi \right)}
\end{equation}
with differential
\begin{equation}
    \dif t = \frac{2}{\ln\left( 1+ k_\text{max}/\xi \right)} \dif
    \left[ \ln\left( 1 + k/\xi\right) \right].
\end{equation}
Therefore we rewrite Eq.~(\ref{eq:int_cov2}) as a function of $t$:
\begin{equation}
    \label{eq:int_cov3}
    \begin{aligned}
    \Delta F^{\text{(CM)}} = \int_{-1}^1 & \biggl \{ \dif t \ln\left( 1+
    \frac{k_\text{max}}{\xi} \right) \\ & \left[k(t)+\xi \right]
    \frac{1}{4} \left \langle |\mathbf{r}-\mathbf{r}_0|^2 \right
    \rangle_{k}^\text{(CM)} \biggl \},
    \end{aligned}
\end{equation} 
where $k(t)$ can be found by rearranging Eq.~(\ref{eq:t}).
An example of the variable transform is shown in Fig.~\ref{fig::u2_vs_k}.

It is straightforward to perform GL quadrature for a general number of
abscissas $n \geq 2$ \cite{weisstein_lobatto_2014}, because
\begin{equation}
    \label{eqn::gl_integration_general_form}
    \int_{-1}^1 \dif t f(t) = w_1 f(-1) + \sum_{i = 2}^{n-1} w_i
    f(t_i) + w_n f(1),
\end{equation}
where $w_i$ are the weights and $t_i$ are the abscissas.  The
abscissas different from $-1, 1$ are the $n-2$ roots of $\dif
P_{n-1}(t)/\dif t$, with $P_{n-1}$ a Legendre polynomial.  We evaluate
this sum numerically using Numpy's Legendre module
\cite{NumpyRef}.  The weights $w_i$ can also be evaluated numerically
for general $n \geq 2$, since they are related to $P_{n-1}$ evaluated
at $t_i$ \cite{weisstein_lobatto_2014}.  For all results in this work,
we choose $n=16$ abscissas.
\section{\label{sec::sampling_th_integrand}Sampling the integrand:
  Hamiltonian Parallel Tempering}
To compute the integral in Eq.~(\ref{eq:int_cov3}), we need to measure
the integrand for different values of $k$, as given by
Eq.~(\ref{eq:t}).  Equilibration of the corresponding simulations can
be accelerated using extensions of the parallel tempering technique,
where replicas differ in chemical potential \cite{Yan99} or in the
potential energy function \cite{Bunker00, Fukunishi02}.

The Parallel-tempering acceptance rule for a swap of configurations between replicas with
different Hamiltonians  follows from the condition of detailed
balance:
\begin{equation}
    \label{eq:ham_det_bal_rem}
    \begin{aligned}
    &\frac{\text{acc}[(\mathbf{r}_i,V_i),(\mathbf{r}_j,V_j)
          \rightarrow
          (\mathbf{r}_j,V_i),(\mathbf{r}_i,V_j)]}{\text{acc}[(\mathbf{r}_i,V_j),(\mathbf{r}_j,V_i)
          \rightarrow (\mathbf{r}_i,V_i),(\mathbf{r}_j,V_j)]} \\ &=
      \frac{\exp \{ -\beta [V_i(\mathbf{r}_j) +
          V_j(\mathbf{r}_i)]\}}{\exp\{ -\beta [V_i(\mathbf{r}_i) +
          V_j(\mathbf{r}_j)]\}} \\ &= \exp \left \{-\beta \left[
        \left(V_i(\mathbf{r}_j)+V_j(\mathbf{r}_i)\right) -
        \left(V_i(\mathbf{r}_i)+V_j(\mathbf{r}_j)\right) \right]
      \right \},
    \end{aligned}
\end{equation}
where $\text{acc}[\cdot\rightarrow\cdot]$ denotes the swap acceptance
probability.  For the particular case of replicas coupled to a
reference state $\mathbf{r}_0$ by a harmonic potential with different
coupling strengths $k_i$, we find the swap acceptance rule
\begin{equation}
  \begin{aligned}
    &\text{acc}[(\mathbf{r}_i,V_i),(\mathbf{r}_j,V_j) \rightarrow
      (\mathbf{r}_j,V_i),(\mathbf{r}_i,V_j)] \\ &= \min \left\{1, \exp
    [\frac{\beta}{2}\left[(k_j-k_i)(|\mathbf{r}_j - \mathbf{r}_0|^2 -
        |\mathbf{r}_i - \mathbf{r}_0|^2)\right] \right\}.
  \end{aligned}
\end{equation}
To check whether the replicas are well equilibrated, we consider the
correlations in the ``time series'' of $|\mathbf{r}-\mathbf{r}_0|_{k}$
\emph{vs} number of Monte Carlo steps for each replica.
\section{\label{sec::Einstein_crystal_CM}Einstein crystal}
The harmonic potential is defined as follows:
\begin{equation}
    V(\mathbf{r}|\mathbf{r}_0,k) = \frac{k}{2}
    |\mathbf{r}-\mathbf{r}_0|^2 = \frac{k}{2} \sum_i^N
    |\mathbf{r}_i-\mathbf{r}_{i,0}|^2
\end{equation}
where $\mathbf{r}_0$ denotes the equilibrium position, the index $i$
denotes the $i$-th of $N$ particles, each with $d$ degrees of freedom,
and we have assumed that the spring constant $k$ is the same for all
directions of motion.  We can compute the mean squared particle
displacement for a harmonic oscillator in the canonical ensemble
analytically.  We start with the partition function:
\begin{equation}
    \label{eq:z_har}
    Z_{NVT} = \left(\frac{2 \pi}{\beta k}\right)^{\frac{dN}{2}}.
\end{equation}
We consider the free energy for the system $F = -\beta^{-1} \ln Z$
and observe that
\begin{equation}
    \label{eq:free_energy_u2_der} \begin{aligned} \left(\frac{\partial
    F(k)}{\partial k}\right)_{NVT} =
    &-\beta^{-1}\frac{\partial}{\partial k} \ln Z = -(\beta
    Z)^{-1}\frac{\partial Z}{\partial k} \ln Z \\ =
    & \frac{\displaystyle\int_{-\infty}^{\infty} \dif \mathbf{r}^{dN} \frac{1}{2}
    |\mathbf{r}-\mathbf{r}_0|^2 e^{-\beta
    k|\mathbf{r}-\mathbf{r}_0|^2/2}}{\displaystyle\int_{-\infty}^{\infty} \dif \mathbf{r}^{dN}
    e^{-\beta k|\mathbf{r}-\mathbf{r}_0|^2/2}}\\ =
    & \frac{1}{2} \langle |\mathbf{r}
    - \mathbf{r}_0|^2\rangle, \end{aligned}
\end{equation}
hence we can compute the mean squared distance for a $dN$-dimensional
harmonic oscillator 
\begin{equation}
    \label{eq:free_energy_u2}
    \langle |\mathbf{r} - \mathbf{r}_0|^2\rangle = 2
    \left(\frac{\partial F(k)}{\partial k}\right)_{NVT} =
    \frac{dN}{\beta k}.
\end{equation}
For thermodynamic integration we are interested in the limit
$k\rightarrow 0$.  In this limit there is no penalty for moving the
system as whole, hence the mean squared displacement becomes of the
order of $L^2$, where $L$ is the box side length.  This result means
that the function $\langle |\mathbf{r} - \mathbf{r}_0|^2\rangle_k$
will be strongly peaked at $k=0$, thus making its integration
difficult.  For this reason, we would like this function to vary
slowly with $k$.  This behaviour can be achieved by fixing the centre
of mass of the system, so that drifting as a whole is
forbidden \cite{Frenkel02}.

The centre of mass is defined as:
\begin{equation}
    \mathbf{r}_\text{CM} = \sum_i \mu_i \mathbf{r}_i, ~~
    \text{where}~\mu_i=\frac{m_i}{\sum_im_i}.
\end{equation} 
When computing the potential energy for the harmonic spring, we must
apply the following correction:
\begin{equation}
    |\mathbf{r}^{(C)} - \mathbf{r}_0^{(C)}|^2 = \sum_i^N
    |\mathbf{r}_i^{(U)} - \mathbf{r}_{i,0}^{(U)} - \Delta
    \mathbf{r}^\text{(CM)}_i|^2,
\end{equation}
where $i$ is the index for the $i$-th particle and $C$ and $U$ denote
the corrected and the uncorrected coordinates respectively. The
configurational partition function requires a correction, hence we
define the corrected partition function $Z_\text{CM}$ with centre of mass
fixed at $\mathbf{r}_\text{CM}=\mathbf{0}$ and note that:
\begin{equation}
    \label{eq:har_zcm} \begin{aligned} Z_\text{CM}=
    &\int_{-\infty}^{\infty} \dif \mathbf{r}^{dN} e^{-\beta
    k|\mathbf{r}-\mathbf{r}_0|^2/2}\delta\left(\sum_i\mu_i\mathbf{r_i}\right)\\
    =&\left(\frac{\beta
    k}{2\pi\sum_i\mu_i^2}\right)^{d/2}\left(\frac{2\pi}{\beta
    k}\right)^{Nd/2} = \left(\frac{\beta
    k}{2\pi\sum_i\mu_i^2}\right)^{d/2} Z, \end{aligned}
\end{equation}
where solution of the integral was obtained after a fair amount of
algebra by rewriting the Dirac delta as the Fourier sum
$\delta(\mathbf{x}) = 1/(2\pi^3)\int \dif \mathbf{k}
\exp(i\mathbf{k}\mathbf{x})$ \cite{Polson00, Vega08}.

Using Eq.~(\ref{eq:free_energy_u2}) we find the mean squared
displacement for the constrained Harmonic oscillator:
\begin{equation}
    \label{eq:free_energy_u2_cm}
    \langle |\mathbf{r} - \mathbf{r}_0|^2\rangle_\text{CM} = 2
    \left(\frac{\partial F^\text{CM}(k)}{\partial k}\right)_{NVT} =
    \frac{(N-1)d}{\beta k}.
\end{equation}
This result can be interpreted as the mean squared displacement of the
$(N-1)d$ harmonic oscillator: fixing the centre of mass is equivalent
to fixing one particle and integrating Eq.~(\ref{eq:har_zcm}) over the
remaining degrees of freedom by doing the change of variables
$\mathbf{r}'_i=\mathbf{r}_i-\mathbf{r}_N$ (conveniently with unit
Jacobian) if the $N$-th particle is fixed.

To conclude, let us relabel the potential as
\begin{equation}
    V(\mathbf{r}|\mathbf{r}_0,k,\lambda) = (1-\lambda)
    \Phi(\mathbf{r}) + \frac{1}{2} \lambda k
    |\mathbf{r}-\mathbf{r}_0|^2,
\end{equation}
where $\Phi(\mathbf{r})$ is an arbitrary field, it could be, for
instance, an additional inter-atomic interaction independent of $k$
or, even the zero field.  Let us consider the limit
$\lambda \rightarrow 0$: from the ratio of the partition functions for
the constrained and unconstrained centre of mass, we find:
\begin{equation}
    \begin{aligned} \frac{Z_\text{CM}(\lambda=0)}{Z(\lambda=0)} =
    &\frac{\displaystyle\int_{-\infty}^{\infty} \dif \mathbf{r}^{dN}
    e^{-\beta \Phi(\mathbf{r})} \delta(\sum_i\mu_i\mathbf{r_i})}{\displaystyle\int_{-\infty}^{\infty} \dif \mathbf{r}^{dN}
    e^{-\beta \Phi(\mathbf{r})}}\\ =
    &\left \langle \delta\left(\sum_i \mu_i \mathbf{r}_i \right)\right \rangle
    = \mathcal{P}(\mathbf{r}_\text{CM}=\mathbf{0}), 
    \end{aligned}
\end{equation}
where $\delta$ is the Dirac delta function and
$\mathcal{P}(\mathbf{r}_\text{CM}=0)$ is the probability density of
the centre of mass being at $\mathbf{\mathbf{0}}$ when $\lambda=0$.
Hence we write:
\begin{equation}
    \label{eq:unc_zcm}
    Z_\text{CM}(\lambda=0) = Z(\lambda=0) \mathcal{P}(\mathbf{r}_\text{CM}=0)
\end{equation}
where $\mathcal{P}$ depends on the details of the system.  If the
equilibrium structure is invariant to translations, a condition that
holds true in a system with periodic boundary conditions, then we can
take $\mathcal{P}$ = $1/V_\text{cell}$, where $V_\text{cell}$ is the
smallest repeating unit in the periodic system (unit cell).  This is
at worst $V_\text{cell}=V_\text{box}$, while for a fcc Einstein
crystal it would correspond to the Wigner-Seitz cell $V_{cell} =
V_{box}/N$
\cite{Frenkel84}.
\section{\label{sec::poly_fluid}Polydisperse hard-sphere fluid and
  total accessible volume}
We can write the total accessible volume as

\begin{equation}
    -\ln\mathcal{V}(N,\phi) = -N\ln V_\text{box} +
    Nf_{\text{ex}}(\phi), 
\end{equation}

where $\phi$ is the volume fraction and $f_{\mathrm{ex}(\phi)}$ is the
excess free energy, which is the difference in free energy between the
hard sphere fluid and the ideal gas.  We can compute the excess free
energy by thermodynamic integration \cite{Frenkel02}.  We start by
noting that $\partial F /\partial(1/V_\text{box}) = V_\text{box}^2 P$
and define the number density $\rho = N/V_\text{box}$, hence we write
\begin{equation}
    f_{\text{ex}}(\rho) = \frac{F(\rho)}{N} -
    \frac{F^{(\text{id})}(\rho)}{N} = \int_0^{\rho} \dif \rho' \left (
    \frac{P(\rho')-\rho'}{\rho'^2} \right).
\end{equation}
By noting that the volume fraction of a polydisperse system is $\phi =
v_d \rho \langle \sigma^d \rangle$ \cite{Santos99}, where $v_d$ is the
volume of the $d$-dimensional unit sphere and $\langle \sigma^d
\rangle$ is the $d$-th moment of the distribution of diameters, we can
change variable and write

\begin{equation}
    \label{eq:excess_f}
    f_{\text{ex}}(\phi) = \frac{F(\phi)}{N} -
    \frac{F^{(\text{id})}(\phi)}{N} = \int_0^{\phi} \dif \phi' \left (
    \frac{Z(\phi')-1}{\phi'} \right),
\end{equation}

where $Z(\phi)= P/\rho$ is the compressibility factor (we set $\beta
=1$ everywhere).

Analytical approximations for the compressibility factors for the two
and three-dimensional polydisperse hard sphere fluid have been
proposed.  For the hard disk fluid we use the Santos-Yuste-Haro (eSYH)
equation of state \cite{Santos99}

\begin{equation}
    \label{eq:shy}
    \begin{aligned}
    Z^\text{poly}_\text{eSHY}(\phi) &= \frac{\langle \sigma
      \rangle^2/\langle \sigma^2 \rangle}{1-2\phi + (2\phi_0 -
      1)\phi^2/\phi_0^2} \\ &+ \frac{1}{1-\phi}\left( 1-\frac{\langle
      \sigma \rangle^2}{\langle \sigma^2 \rangle}\right),
    \end{aligned}
\end{equation} 

where $\phi_0 = \pi/\sqrt{12}$ is the crystalline close packing
fraction.

For three-dimensional fluids, depending on the volume fraction, we
choose two different equations of state.  For volume fraction
$\phi>0.5$, Santos \emph{et al.} \cite{Santos99} suggest the following
equation of state based on the Carnahan-Startling (CS) equation of
state for the monodisperse fluid:

\begin{equation}
    \begin{aligned}
    \label{eq:cs}
    Z^\text{poly}_\text{eCS}(\phi) = 1 + &\left[
      \frac{1+\phi+\phi^2-\phi^3}{(1-\phi)^3} - 1\right] \\ &
    \times \frac{\langle \sigma^2 \rangle}{2 \langle \sigma^3 \rangle^2}
    \left( \langle \sigma^2 \rangle^2 + \langle \sigma \rangle \langle
    \sigma^3 \rangle \right)\\ &+ \frac{\phi}{1-\phi} \left[ 1-
      \frac{\langle \sigma^2 \rangle}{\langle \sigma^3 \rangle}^2
      \left( 2\langle \sigma^2 \rangle^2 - \langle \sigma \rangle
      \langle \sigma^3 \rangle \right) \right].
    \end{aligned}
\end{equation} 

For volume fractions $\phi \leq 0.5$ the eCSK equation of state should
be preferred (based on the Carnahan-Starling-Kolafa equation of state
for the monodisperse fluid)

\begin{equation}
    \label{eq:csk}
    \begin{aligned}
    Z^\text{poly}_\text{eCSK}(\phi) &= Z^\text{poly}_\text{eCS}(\phi)
    \\ &+ \frac{\phi^3(1-2\phi)}{(1-\phi)^3} \frac{\langle \sigma^2
      \rangle}{6\langle \sigma^3 \rangle^2}\left( \langle \sigma^2
    \rangle^2 + \langle \sigma \rangle \langle \sigma^3
    \rangle\right).
    \end{aligned}
\end{equation}

The excess free energy can thus be obtained by substituting one of
Eq.~(\ref{eq:shy}) to (\ref{eq:csk}) in the integral of
Eq.~(\ref{eq:excess_f}), which can then be evaluated numerically for
the desired volume fraction.

\section{\label{sec::gen_edwards} Thermodynamic limit of the
generalised configurational entropy}

From the fit to the empirical c.d.f. of $\mathcal{B}(\mathcal{P})$
with the generalised log-normal cumulative distribution function, corresponding to
Eq.~(\ref{eq:gen_lognormal}), we obtain the set of parameters $\mu$,
$\sigma$, and $\zeta$. From the inset in Fig.~\ref{fig::gen_fits}, we
observe that the mean $\mu$ and scale parameter $\sigma$ scale
linearly with $1/N$. In particular we note that $\sigma$ seems to
approach zero in the thermodynamic limit, as expected. Furthermore we
note that the shape parameter $\zeta$ seems to be approximately
independent of $1/N$ and to have a value of approximately $2$ for all
system sizes, thus suggesting that the distributions of pressures are
consistent with a log-normal distribution.

Therefore, under the reasonable assumption that the biased
distribution of pressures $\mathcal{B}(x|V)$ is log-normal, we write
the integrand in Eq.~(\ref{eq:gen_edw_entropy}) as
\begin{equation}
\begin{aligned}
& I(x; \mu,\sigma, N) \equiv ~\mathcal{B}(x|V)x^{N/\kappa} =
  \\ &
  \frac{1}{x\sqrt{2\pi\sigma^2}}\exp\left(-\frac{(\ln(x)-\mu)^2}{2\sigma^2}+\frac{N}{\kappa}\ln(x)\right),
\end{aligned}
\end{equation}
which is a unimodal distribution with mode $x_{M} =
\exp(N\sigma^2/\kappa + \mu - \sigma^2)$. The distribution
is such that
\begin{equation}
\label{eq:tl_int}
\int_0^{\infty}I(x;\mu,\sigma, N)\dif x =
\exp\left[\frac{\sigma^2N^2}{2\kappa^2}+\frac{N\mu}{\kappa}\right].
\end{equation}
Since $\sigma \propto 1/N$, for large $N$ we have
$\int_0^{\infty}I(x;\mu,\sigma, N \gg \kappa)\dif x = \varepsilon \exp(N
\mu / \kappa)$, with $\varepsilon$ some constant. Thus in the
thermodynamic limit ($N,V, 1/\sigma \to \infty$) we obtain the
expression for the Gibbs configurational entropy per particle, see
Eq.~(\ref{eq:entropy_per_particle2}).

Consider Eq.~(\ref{eq:pressures_f_law}) which we rewrite in terms of
the dimensionless free energy per particle
\begin{equation}
\label{eq:pressures_f_law_tl}
f(\mathcal{P} | \phi_\text{SS}) \equiv -\frac{1}{N}\ln(v) = c
+ \frac{1}{\kappa}\ln(\mathcal{P}),
\end{equation}
where $c=\mathcal{C}(N)/N$; to test the consistency of the results
thus obtained we compare the expression for Edwards configurational
entropy, see Eq.~(\ref{eq:entropy_per_particle2}), to the Gibbs
configurational entropy per particle, see
Eq.~(\ref{eq:entropy_per_particle1}). We thus find that
\begin{equation}
\langle f \rangle_{\mathcal{B}}= c +
\frac{1}{\kappa}\langle \ln (\mathcal{P}) \rangle_{\mathcal{B}},
\end{equation}
which is consistent with Eq.~(\ref{eq:pressures_f_law_tl}) in the
thermodynamic limit. We have thus correctly recovered the power law
relation between pressure and basin volume,
Eq.~(\ref{eq:pressures_f_law}).
 
\bibliography{basinvolume_method}
\end{document}